\documentclass[aps,pra,reprint,amsmath,amssymb,longbibliography,floatfix,nofootinbib]{revtex4-2}
\usepackage[T1]{fontenc}
\usepackage{fix-cm}
\usepackage{amsmath,amssymb,amsthm,bm,graphicx,booktabs}
\usepackage{newtxtext,newtxmath}
\usepackage[colorlinks=true,allcolors=blue]{hyperref}
\hypersetup{pdftitle={Boundary-rank obstructions and measurement-assisted recovery in dissipative flat-band preparation},pdfauthor={Mingdi Xu}}
\graphicspath{{figures/}}
\newcommand{\D}{\mathcal D}
\newcommand{\Lcal}{\mathcal L}
\newcommand{\Qcal}{\mathcal Q}
\newcommand{\Ecal}{\mathcal E}
\newcommand{\Acal}{\mathcal A}
\newcommand{\PF}{P_F}
\newcommand{\Hb}{\mathcal H_b}
\newcommand{\Hf}{\mathcal H_f}
\DeclareMathOperator{\Tr}{Tr}
\DeclareMathOperator{\rank}{rank}

\DeclareMathOperator{\Dest}{Dest}
\DeclareMathOperator{\spec}{spec}
\newtheorem{theorem}{Theorem}
\newcommand{\figfile}[2]{\includegraphics[width=#2]{#1}}
\begin{document}
\title{Boundary-rank obstructions and measurement-assisted recovery in dissipative flat-band preparation}
\author{Mingdi Xu}
\email{2120230195@mail.nankai.edu.cn}
\affiliation{School of Physics, Nankai University, Tianjin 300071, China}
\date{September 29, 2026}
\begin{abstract}
Number-conserving cooling can fail to prepare an interacting flat-band target when Pauli exclusion blocks its transfer destinations. We study this failure on vertex-edge decorated graphs at one particle per flat orbital. The rank of a cut block of the flat-orbital Gram matrix bounds the number of residual source directions that can remain flat. For sublinear-range cooling with no Hamiltonian term, separated spin domains on periodic decorated hypercubic lattices support exponentially many stationary states with a nonzero bright-particle density. For a chain with fixed finite cooling range, we construct a physical Fock initial state whose overlap with a wrong dark state is independent of system size. This gives fidelity and bright-density bounds valid at every time. In higher dimensions, the bare overlap decays with boundary area, while local boundary rotations prepare states with a finite failure weight in constant circuit depth. Dephasing the occupations of all compact bright modes restores global attraction to the ferromagnetic target when the graph and cooling destinations satisfy the stated conditions. The result allows Hamiltonians that preserve the target. The compressed dynamics in the strong-dephasing limit also remains attractive and has a positive gap at each fixed size. Gram calculations, finite-system dynamics, and Liouvillian spectra test the analytic results. Weak-dephasing and particle-transport bounds constrain the preparation time despite eventual convergence.
\end{abstract}
\maketitle

\section{Introduction}

Reservoir engineering prepares a quantum state by designing dissipative processes that drive the system toward it~\cite{Poyatos1996Reservoir,Diehl2008Reservoir,Kraus2008Preparation,Verstraete2009Dissipation}. Engineered dissipative dynamics and entangled steady states have been realized with trapped ions~\cite{Barreiro2011Simulator,Lin2013Dissipative} and superconducting circuits~\cite{Shankar2013Autonomous}. For fermions, proposals address pairing~\cite{Diehl2010Dwave,Yi2012Pairing}, antiferromagnetic order~\cite{Kaczmarczyk2016Antiferromagnet}, and number-conserving topological superconductivity~\cite{Iemini2016NumberConserving}. These examples motivate identifying the restrictions of a particular protocol. In number-conserving interband cooling, Pauli exclusion can block a transfer while its source remains excited. The intended target may therefore be dark without attracting every initial state.

Number-conserving interband cooling has been studied in optical-lattice systems coupled to a phonon reservoir~\cite{Griessner2006Immersion,Griessner2007Cooling}. More recent analyses of two-band fermionic protocols derive reaction-diffusion dynamics and diffusive modes~\cite{Nosov2023Reaction,Lyublinskaya2023Diffusive}, and predict instability of the engineered dark state in the regimes they consider~\cite{Lyublinskaya2025Instability}. Pokart \emph{et al.} explicitly constructed a second dark state in a restricted cooling protocol. They also proved uniqueness after adding dissipative channels, with a jump-sequence argument that excludes mixed stationary states~\cite{Pokart2026Diffusion}. These works already show that specific number-conserving protocols can suffer blocked cooling and that additional channels can restore uniqueness. The question here is how compact-orbital geometry controls the size, excitation content, and initial-state weight of the blocked space.

Flat bands admit real-space descriptions in terms of compact localized orbitals~\cite{Leykam2018FlatBandReview}. Their interacting ferromagnets provide a setting in which the obstruction can be related to spatial geometry. Line-graph Hubbard models establish the role of degenerate one-particle ground states in ferromagnetism~\cite{Mielke1991GroundStates,Mielke1991Further}. In decorated-lattice constructions, overlapping compact orbitals and repulsive on-site interactions select maximal spin through connectivity~\cite{Tasaki1992Ferromagnetism,Tasaki1998Review}. The same compact support that makes local cooling possible also allows one spin species to fill all destinations in a region. Two spatially separated spin regions avoid doublons and can remain invisible to a local spin-selection process. A count of the occupied source coordinates is insufficient to quantify this failure: the compact orbitals within each band are not orthogonal, and a source coordinate generally has both flat and bright components.

We characterize the obstruction using the rank of the Gram matrix across a spatial cut. Once all flat orbitals in a region are filled, the cut rank bounds the number of independent residual source directions that project into the unfilled flat space. The remaining source particles must occupy the physical bright subspace. The resulting spectral-support bound holds throughout the constructed many-body space, including superpositions of its nonorthogonal exterior-basis vectors. On periodic decorated hypercubic graphs, that space has a positive entropy density and an extensive bright population.

A large dark space need not have appreciable overlap with a readily prepared state. We therefore construct the initial states explicitly. On a chain, a physical Fock state loses overlap with a dark witness only at a fixed number of boundary modes. Its fidelity deficit and bright density remain bounded away from zero for all times. In higher dimensions, the bare overlap decreases with boundary area. Local rotations on the boundary instead prepare a specified family with a size-independent failure weight in constant depth. Typical random or thermal states require a separate analysis.

The same model admits a local repair. Occupation measurements of overlapping compact bright modes generate the algebra of bright-mode rearrangements. Together with cooling and doublon selection, this excludes every wrong invariant subspace and establishes convergence for mixed initial states. Any Hamiltonian that preserves the particle sector and has the target as an eigenstate can be included. Using the established strong-dissipation projection~\cite{Burgarth2019StrongCoupling}, we then show that the compressed generator remains attractive. This is a finite-size convergence statement; preparation times still depend on the measurement strength and on transport across the system.

\begin{figure*}[t]
\centering
\figfile{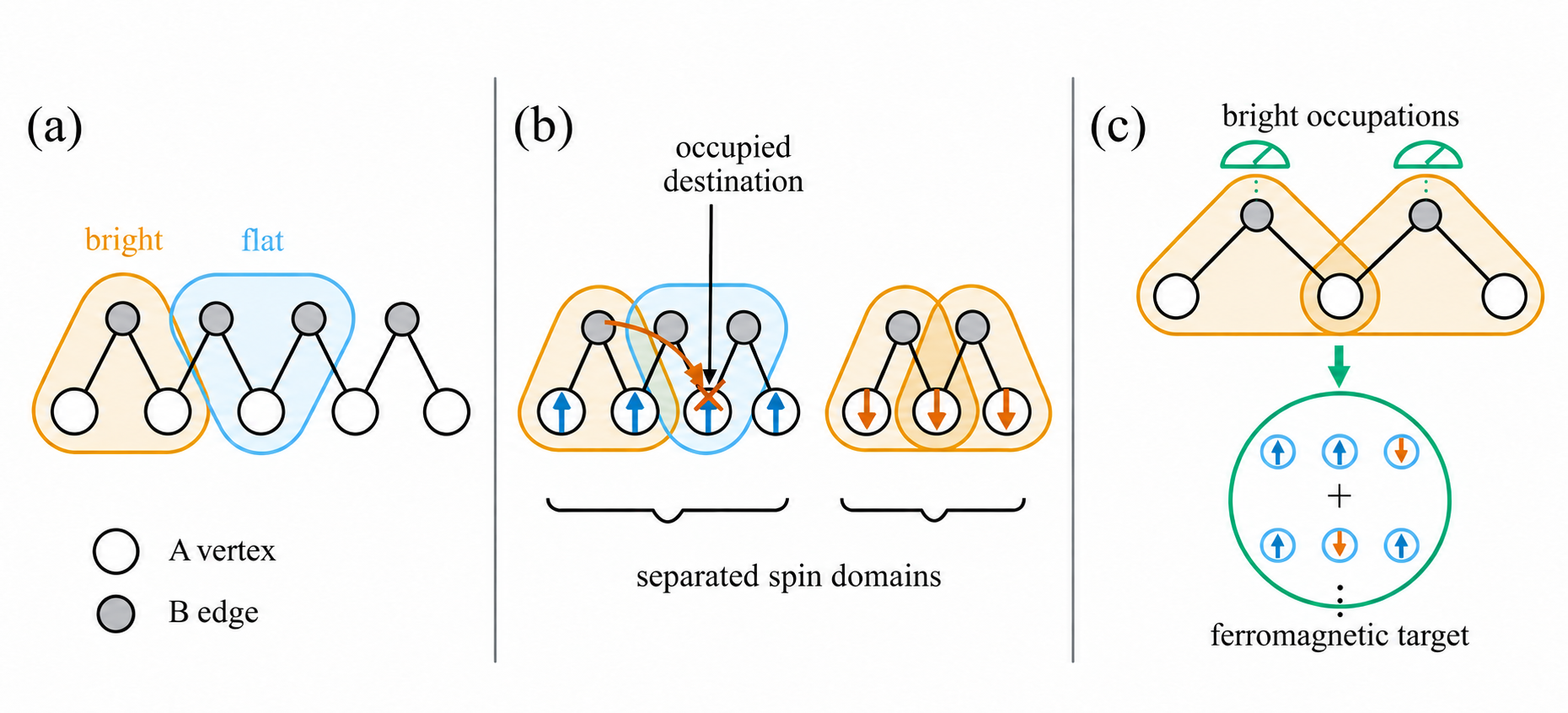}{17.8cm}
\caption{Geometry and dissipative operations. (a) A vertex-edge decorated chain illustrates the compact supports: a flat orbital is centered on an $A$ vertex and overlaps its incident $B$ decorations, whereas a bright orbital is centered on a $B$ decoration and overlaps its two endpoint $A$ sites. Colored contours indicate support, not amplitudes or mutually orthogonal orbitals. (b) Filling all flat destinations of occupied sources within separated spin domains blocks transfer while avoiding physical doublons. The diagram is conceptual; Eqs.~\eqref{eq:regions} and \eqref{eq:chain-initial} specify the actual many-body constructions. (c) Adjacent bright-occupation measurements overlap on an $A$ site and do not commute. With the original cooling and selectors still present, they restore attraction to the coherent fixed-magnetization ferromagnet. The displayed spin words represent terms in that superposition, not a classical spin product.}
\label{fig:model}
\end{figure*}

\section{Decorated graph and preparation protocol}
\label{sec:model}

Let $G=(V,E)$ be a finite connected simple graph with $M=|V|$ vertices and $B=|E|$ edges. The physical sites are $A_v$ at each vertex and $B_e$ on each edge. For $e=(v,w)$ and $\lambda>0$, define the unnormalized orbitals
\begin{align}
 f_v^\dagger&=c_{A_v}^\dagger-\lambda\sum_{e\ni v}c_{B_e}^\dagger,\nonumber\\
 d_e^\dagger&=c_{B_e}^\dagger+\lambda(c_{A_v}^\dagger+c_{A_w}^\dagger),
 &u_e^\dagger=c_{B_e}^\dagger.
\label{eq:orbitals}
\end{align}
Spin indices are restored when needed. With $\mathsf B$ the unsigned vertex-edge incidence matrix, the column frames are
\begin{equation}
 \mathsf F=\binom{I_M}{-\lambda\mathsf B^{\mathsf T}},\qquad
 \mathsf D=\binom{\lambda\mathsf B}{I_B}.
\end{equation}
They obey $\mathsf F^\dagger\mathsf D=0$. The flat and bright one-particle spaces $\Hf$ and $\Hb$ have dimensions $M$ and $B$, but each compact frame is internally nonorthogonal. In particular,
\begin{equation}
 G_f=I_M+\lambda^2\mathsf B\mathsf B^{\mathsf T},\quad
 \Pi_f=\mathsf F G_f^{-1}\mathsf F^\dagger,\quad \Pi_b=I-\Pi_f.
\label{eq:projectors}
\end{equation}
Compact support does not imply mutual orthogonality. Real-space constructions of flat bands exploit compact localized states~\cite{Flach2014Detangling,Maimaiti2017Generators}; their relation to orthogonal band projectors and the projectors' spatial decay has been analyzed explicitly~\cite{Kim2026Projectors}. Here Eq.~\eqref{eq:projectors} fixes the physical subspaces used throughout. The physical bright-number operator is
\begin{equation}
 N_b=\sum_\sigma c_\sigma^\dagger\Pi_b c_\sigma.
\label{eq:bright-number}
\end{equation}
It is not the occupation count of the $u_e$ coordinates.

We work in the fixed sector $N_\uparrow=p$, $N_\downarrow=q$, with $p+q=M$ and $p,q\geq1$. This is one particle per flat orbital, not half filling of all $M+B$ physical sites. The target $|F_m\rangle$, $m=(p-q)/2$, is the fixed-magnetization component of the saturated ferromagnet obtained by filling the flat band with one spin and applying the physical global spin-lowering operator
$S^-=\sum_{v\in V}c_{A_v\downarrow}^\dagger c_{A_v\uparrow}
+\sum_{e\in E}c_{B_e\downarrow}^\dagger c_{B_e\uparrow}$.
Write $\PF=|F_m\rangle\langle F_m|$ and $F(t)=\Tr[\PF\rho(t)]$.

For each source edge, choose a destination set $\Dest(e)\subset V$. The independent cooling channels are
\begin{equation}
 C_{ev\sigma}=f_{v\sigma}^\dagger d_{e\sigma},\qquad v\in\Dest(e).
\label{eq:cooling}
\end{equation}
All directed incidence links $x\to y$ carry the selector
\begin{align}
 P_{x\to y}&=T_{xy}^\dagger c_{x\downarrow}c_{x\uparrow},\nonumber\\
 T_{xy}^\dagger&=\frac{c_{x\uparrow}^\dagger c_{y\downarrow}^\dagger+
 c_{x\downarrow}^\dagger c_{y\uparrow}^\dagger}{\sqrt2}.
\label{eq:selector}
\end{align}
The two directions are both included. We use the Lindblad convention~\cite{Gorini1976Generators,Lindblad1976Generators}
\begin{equation}
 \D[J]\rho=J\rho J^\dagger-\tfrac12\{J^\dagger J,\rho\}.
\end{equation}
The generator $\Lcal_0$ is the positive-rate sum of the cooling and selector dissipators, with actual Hamiltonian $H=0$. Rescaling a jump by a nonzero constant preserves its kernel but changes the dynamics. We therefore specify jump normalizations and rates for each numerical comparison.

The common selector kernel is exactly the physical no-doublon space in the sector considered here. Indeed, with $q_x=n_{x\uparrow}n_{x\downarrow}$,
\begin{equation}
 \|P_{x\to y}\psi\|^2=
 \langle q_x[1-(n_{y\uparrow}+n_{y\downarrow})/2]\rangle_\psi.
\end{equation}
Vanishing in both directions implies $q_x\psi=q_y\psi$. Connectivity propagates this equality throughout the physical graph; a state with only $M$ particles cannot doubly occupy every physical site. Within the flat band, no doublons at $A_v$ force one particle per flat label. The constraints at $B_e$ identify the coefficients of spin words related by an endpoint exchange. Graph connectivity then selects $|F_m\rangle$ uniquely, as in the flat-band ferromagnetism argument~\cite{MielkeTasaki1993,Tasaki1998Review}. Details and conventions are in Supplemental Material Sec.~S1.

A useful parent Hamiltonian is
\begin{equation}
 H_{\rm par}=t_0\sum_{e\sigma}d_{e\sigma}^\dagger d_{e\sigma}
 +U\sum_xq_x,\qquad t_0,U>0.
\label{eq:parent}
\end{equation}
It annihilates the target and satisfies $H_{\rm par}\geq t_0N_b$, so the bright-number bound also gives an energy bound. We use this Hamiltonian as a diagnostic in the obstruction problem and allow it in the actual dynamics of the recovery protocol below.

\section{A geometric obstruction to cooling}
\label{sec:obstruction}

\subsection{The complete independent-transfer kernel}

General Lindblad dynamics can retain initial-state information through its nondecaying subspaces and conserved quantities~\cite{Baumgartner2008General,Albert2014Symmetries}. We construct such an obstruction directly from occupied transfer destinations; no additional strong symmetry is assumed.

For one spin, the vectors $\{f_v,u_e\}$ form an invertible, nonorthogonal one-particle coordinate system. Thus $|I,J\rangle=f_I^\dagger u_J^\dagger|0\rangle$, with $|I|+|J|=n$, are an exterior basis. Define $D(J)=\bigcup_{e\in J}\Dest(e)$. The common kernel of all independent cooling channels is exactly
\begin{equation}
 \mathcal K_n=\operatorname{span}\{\,|I,J\rangle:D(J)\subset I\,\}.
\label{eq:kernel}
\end{equation}
To prove necessity, fix a channel $(e,v)$. A nonzero input requires $e\in J$ and $v\notin I$, and its output labels are $(I\cup\{v\},J\setminus\{e\})$. These labels determine the input uniquely, so distinct nonzero exterior-basis outputs cannot cancel. The kernel dimension is
\begin{equation}
 \dim\mathcal K_n=\sum_{J\subset E}
 \binom{M-|D(J)|}{n-|J|-|D(J)|},
\end{equation}
with out-of-range binomial coefficients equal to zero. The full dark space also satisfies the selector constraints. Coherently summing destinations within one channel can enlarge the cooling kernel through cancellation; every vector dark for each independent channel remains dark for such a sum (Supplemental Material Sec.~S2).

\subsection{Cut rank bounds physical bright occupation}

Choose two disjoint vertex regions $U_\uparrow,U_\downarrow$ with no graph edge between them. For a region $U$, let
\begin{equation}
 T(U)=\{e=(v,w):v,w\in U,\ \Dest(e)\subset U\}.
\end{equation}
Fill every flat orbital in each region and $a_\sigma$ legal source coordinates, where $0\leq a_\sigma\leq|T(U_\sigma)|$ and
\begin{equation}
 |U_\uparrow|+a_\uparrow=p,\qquad
 |U_\downarrow|+a_\downarrow=q.
\label{eq:regions}
\end{equation}
The span $W$ of the resulting $f_{U_\uparrow}^\dagger u_{J_\uparrow}^\dagger
f_{U_\downarrow}^\dagger u_{J_\downarrow}^\dagger|0\rangle$, with $J_\sigma\subset T(U_\sigma)$ and $|J_\sigma|=a_\sigma$, has dimension
\begin{equation}
 \dim W=\prod_{\sigma}\binom{|T(U_\sigma)|}{a_\sigma}.
\label{eq:dark-dimension}
\end{equation}
Every cooling destination is occupied. The two physical spin supports are disjoint, so every selector also vanishes.

To locate $W$ in the physical bright-number spectrum, define
\begin{equation}
 r(U)=\rank(G_f)_{U,U^c}\leq|\partial U|,
\label{eq:cutrank}
\end{equation}
where $\partial U$ consists of vertices inside $U$ adjacent to its complement. Let $P_U$ project orthogonally onto the filled flat span and let $R_U=(I-P_U)\operatorname{span}\{u_e:e\in T(U)\}$. Then
\begin{equation}
 \rank(\Pi_f|_{R_U})\leq r(U).
\label{eq:rank-lemma}
\end{equation}
To see the bound, write $x=\mathsf F c\in\Hf$ orthogonal to the filled flat span. Its coefficients obey
\begin{equation}
 c_U=-(G_f)_{U,U}^{-1}(G_f)_{U,U^c}c_{U^c}.
\end{equation}
For an internal edge, $\langle x,u_{(v,w)}\rangle=-\lambda(c_v^*+c_w^*)$ only reads $c_U$. The residual flat overlap therefore depends on at most $r(U)$ independent boundary combinations.

In an $a_\sigma$-particle exterior power, at most $r(U_\sigma)$ particles can occupy residual directions with a flat component. This gives the following statement for every vector in $W$.

\begin{theorem}[Boundary-rank obstruction]
Under the assumptions above,
\begin{equation}
 W\subset\mathbf1_{[h_G,a_\uparrow+a_\downarrow]}(N_b)\mathcal H,
 \quad h_G=\sum_\sigma[a_\sigma-r(U_\sigma)]_+.
\label{eq:bright-support}
\end{equation}
If $h_G>0$, $W$ is orthogonal to the target, and every density matrix supported in $W$ is stationary under $\Lcal_0$.
\end{theorem}

Equation~\eqref{eq:bright-support} concerns spectral support. It survives superpositions of the nonorthogonal exterior-basis vectors and implies $N_b\geq h_GP_W$, where $P_W$ is the physical orthogonal projector onto $W$. Supplemental Material Sec.~S4 gives the full exterior-power argument.

\subsection{Extensive dark spaces at sublinear cooling range}

Take the periodic $d$-dimensional hypercubic base graph of side $n$, so $M=n^d$ and $B=dM$. Let $p/M\to\nu\in(0,1)$, and choose a fixed $0<\theta<d/(d+1)$. Two separated macroscopic strips can satisfy $|U_\uparrow|=(1-\theta)p+o(M)$ and $|U_\downarrow|=(1-\theta)q+o(M)$. If all cooling destinations are within graph distance $R_n=o(n)$ of their source, then
\begin{equation}
 |T(U_\sigma)|=d|U_\sigma|-o(M),\qquad r(U_\sigma)=O(n^{d-1}).
\end{equation}
With the source counts fixed by Eq.~\eqref{eq:regions}, the upper and lower spectral bounds squeeze the bright density to $\theta$, uniformly over normalized states supported in $W$. Stirling's formula gives
\begin{align}
 \lim_{M\to\infty}\frac{\log\dim W}{M}&=s_d(\theta)>0,\nonumber\\
 s_d(\theta)&=d(1-\theta)h_{\rm bin}\!\left(\frac{\theta}{d(1-\theta)}\right),
\label{eq:entropy}
\end{align}
where $h_{\rm bin}(x)=-x\log x-(1-x)\log(1-x)$. The entropy uses natural logarithms. Thus cooling with sublinear range and the stated selectors cannot prepare a unique global attractor on this graph family. Whether longer-range transfers suffice, and whether other flat-band geometries admit the same construction, remain separate questions.

\section{Explicit initial states with all-time failure bounds}
\label{sec:initial}

\subsection{A physical Fock state on a periodic chain}

Let the base graph be a periodic chain of length $L$, with edge $j=(j,j+1)$ and $\lambda=1/\sqrt2$. A fixed finite nonempty offset set $S\subset\mathbb Z$ specifies $\Dest(j)=j+S$. Put
\begin{equation}
 \ell_-=-\min(S\cup\{0\}),\quad \ell_+=\max(S\cup\{0\}),\quad
 w=\ell_-+\ell_+.
\end{equation}
For integers $a,b\geq3$, choose $p=2a+w$, $q=2b+w$, and $L=2(a+b+w)$. Set $m_\uparrow=a+w$, $m_\downarrow=b+w$, and
\begin{align}
 I_\uparrow&=[1,m_\uparrow],\nonumber\\
 I_\downarrow&=[m_\uparrow+2,m_\uparrow+m_\downarrow+1],\nonumber\\
 J_\sigma&=[\min I_\sigma+\ell_-,\max I_\sigma-\ell_+].
\end{align}
The initial state is the normalized physical-site Fock product
\begin{equation}
 |\Phi_S\rangle=\prod_\sigma
 \left(\prod_{i\in I_\sigma}c_{A_i,\sigma}^\dagger\right)
 \left(\prod_{j\in J_\sigma}c_{B_j,\sigma}^\dagger\right)|0\rangle.
\label{eq:chain-initial}
\end{equation}
Products have a fixed fermion ordering. Both occupied spin supports are separated even across the periodic seam.

The corresponding normalized dark vector $|\Psi_S\rangle\propto\prod_\sigma f_{I_\sigma,\sigma}^\dagger u_{J_\sigma,\sigma}^\dagger|0\rangle$ lies entirely in the sector
\begin{equation}
 N_b\geq a+b-4=L/2-w-4.
\end{equation}
For offsets allowing a source at the right boundary of an interval, the source edge need not be internal. The chain proof therefore uses the flat-coefficient recurrence directly, which still leaves at most two residual flat directions per spin interval (Supplemental Material Sec.~S3).

The overlap with Eq.~\eqref{eq:chain-initial} is exactly
\begin{equation}
 c_S=|\langle\Psi_S|\Phi_S\rangle|^2
 =[D_{\ell_-+1}D_{\ell_+}]^{-2},
\label{eq:overlap}
\end{equation}
where $D_0=1$, $D_1=3/2$, and $D_n=2D_{n-1}-D_{n-2}/4$. After the occupied $B$ components are removed from the flat orbitals, the remaining overlap determinant splits into two fixed-size boundary blocks. This is why $c_S$ does not decrease with $L$ for fixed $S$.

For any normalized dark vector $\Psi$, $\Lcal_0^\dagger P_\Psi=\sum_\mu J_\mu^\dagger P_\Psi J_\mu\geq0$. Its weight cannot decrease. Combining this fact with $P_\Psi\perp\PF$ and $N_b\geq(L/2-w-4)P_\Psi$ yields, for every $t\geq0$,
\begin{align}
 F(t;\Phi_S)&\leq1-c_S,\nonumber\\
 \frac{\langle N_b\rangle_t}{L}&\geq c_S\left(\frac12-\frac{w+4}{L}\right).
\label{eq:alltime}
\end{align}
For $S=\{0\}$, $c_S=4/9$ and the limiting bright-density lower bound is $2/9$. For endpoint cooling, $S=\{0,1\}$, these values are $16/81$ and $8/81$. The bounds in Eq.~\eqref{eq:alltime} hold throughout the evolution; the long-time bright density may be larger than its lower bound. The initial state is spatially inhomogeneous. Translations and mixtures of translations obey the same bound, although this construction does not determine the behavior of typical initial states.

\subsection{Boundary cost and boundary preparation in higher dimensions}

For endpoint cooling on a decorated graph, fill all internal source edges $E(U_\sigma)$ along with the flat region $U_\sigma$, with the total particle number constrained to $M$. Compare its dark vector to the physical Fock state occupying all $A_v$ in $U_\sigma$ and all internal $B_e$. If $b_v$ counts edges from $v\in U_\sigma$ to the complement, the exact overlap is
\begin{equation}
 c_{\rm Fock}=\prod_{\sigma}\prod_{v\in\partial U_\sigma}
 (1+\lambda^2b_v)^{-1}.
\label{eq:area}
\end{equation}
Once internal $B$ sites are occupied, the residual flat vector at $v$ is $|A_v\rangle-\lambda\sum_{e\ni v,\,e\in\partial E(U)}|B_e\rangle$. Different residual boundary stars have disjoint support, and their norms give Eq.~\eqref{eq:area}. A growing boundary therefore suppresses the overlap with this particular witness. The bare Fock state's weight in other dark vectors remains undetermined.

A local number-conserving rotation on each boundary star maps the occupied $A_v$ mode into that normalized residual vector. Finite-dimensional mode unitaries admit decompositions into two-mode transformations~\cite{Reck1994Unitary}; for fermionic modes these can be implemented as number-conserving Givens rotations~\cite{Jiang2018FermionAlgorithms}. All stars are disjoint, so bounded graph degree gives constant circuit depth. Apply these rotations except at a fixed set $\mathcal A$ of $k\geq1$ boundary vertices. The resulting state is not already dark, but its dark overlap is
\begin{equation}
 c_{\mathcal A}=\prod_{v\in\mathcal A}(1+\lambda^2b_v)^{-1}
 \geq(1+\lambda^2\Delta)^{-k},
\label{eq:dressed}
\end{equation}
where $\Delta$ is the maximum degree. When $h_G>0$, it obeys $F(t)\leq1-c_{\mathcal A}$ and $\langle N_b\rangle_t/M\geq c_{\mathcal A}h_G/M$ at all times.

For two full transverse strips of widths $m_\uparrow,m_\downarrow$ on a periodic hypercubic graph, the integer condition $n+2=(d+1)(m_\uparrow+m_\downarrow)$ enforces $N=M$. In this sequence, $h_G/M\to d/(d+1)$. Flat boundaries have $b_v=1$, so omitting one gate gives $c_{\mathcal A}=(1+\lambda^2)^{-1}$, whereas the bare overlap is $(1+\lambda^2)^{-4n^{d-1}}$. This preparation uses specified local mode rotations only near the boundary and leaves a fixed number of them unapplied. A finite error probability at every boundary site would require a different bound. The bounded depth concerns only these specified boundary corrections; it does not provide a general preparation procedure for long-range correlations, whose generation is constrained by locality~\cite{Bravyi2006Correlations}.

\begin{figure*}[t]
\centering
\figfile{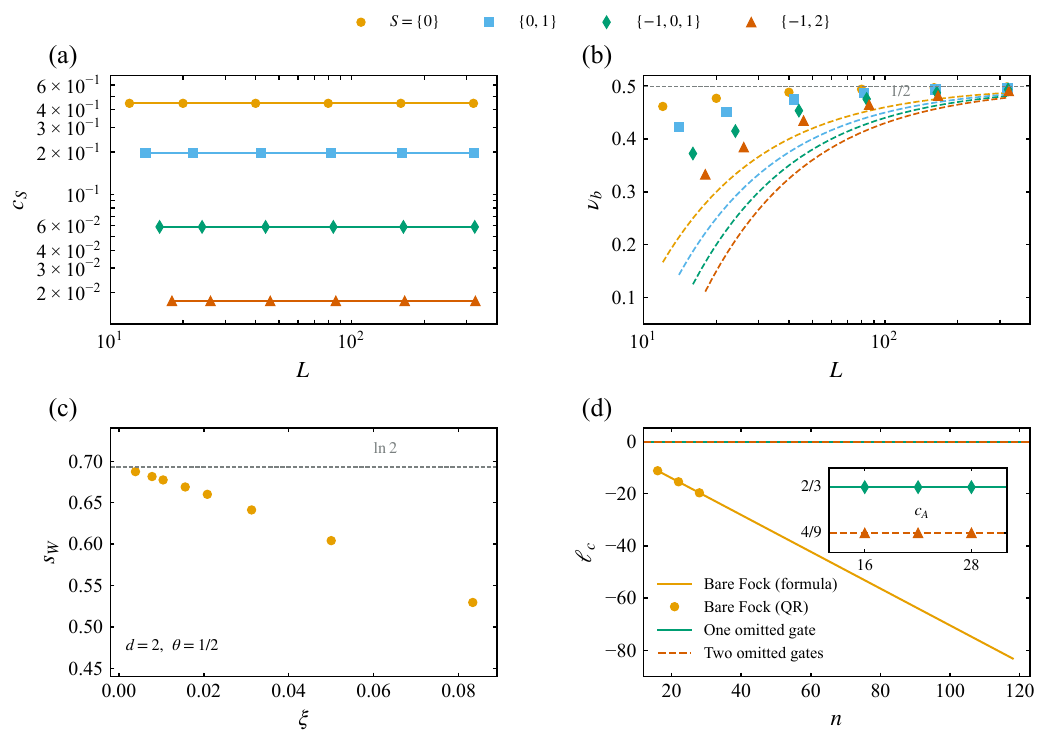}{17.8cm}
\caption{Static checks at $\lambda=1/\sqrt2$. We use $c_S=|\langle\Psi_S|\Phi_S\rangle|^2$, $\nu_b=\langle N_b\rangle_{\Psi_S}/L$, $\xi=1/n$, $s_W=M^{-1}\log\dim W$, and $\ell_c=\log_{10}c_{\rm Fock}$; the inset uses $c_{\mathcal A}$. (a) Physical-orbital QR calculations of $c_S$ (filled points) and the boundary-determinant formula for four fixed exit sets. The plotted sequence uses $a=b=3,5,10,20,40,80$; the data also contain $a=2b$. (b) Physical bright density $\nu_b$ of the corresponding dark Slater states (points), with the spectral lower bound $1/2-(w+4)/L$ (colored dashed curves). The gray line is $1/2$. These are static properties, not dynamical plateaus. (c) Exact combinatorial entropy density $s_W$ of two separated strips on a decorated square lattice at $p=q=M/2$ and $\theta=1/2$, approaching $\log2$ as $\xi\to0$. (d) Bare-Fock overlap for the balanced two-dimensional strip sequence $n=6m-2$: the line is $\ell_c=-4n\log_{10}(3/2)$, with independent QR points at $n=16,22,28$. The inset resolves $c_{\mathcal A}=2/3$ or $4/9$ when one or two boundary gates are omitted; points are independent boundary-star determinant checks. Every plotted construction has $N=M$.}
\label{fig:static}
\end{figure*}

\section{Local measurements restore global attraction}
\label{sec:recovery}

Normalize each compact bright orbital, $\bar d_e=d_e/\|d_e\|$, and add its occupation projector
\begin{equation}
 Q_{e\sigma}=\bar d_{e\sigma}^\dagger\bar d_{e\sigma}.
\end{equation}
The channel $\D[Q_{e\sigma}]$ describes occupation measurement with the outcome discarded. It requires neither postselection nor feedback. Neighboring bright modes overlap, so their occupation projectors generally do not commute.
Operationally, the unread two-outcome measurement of a projector $Q$ is
$\mathcal M_Q(\rho)=Q\rho Q+(I-Q)\rho(I-Q)$, and therefore
$\mathcal M_Q-\mathcal I=2\D[Q]$. Poisson-distributed measurement events at
rate $\kappa/2$ thus generate $\kappa\D[Q]$. This is an ideal operational
interpretation of the Lindblad term, not a claim that a specific experimental
platform already implements the required coherent compact-mode projector.

\begin{theorem}[Measurement-assisted recovery]
Let $G$ be finite, connected, and of minimum degree at least two. Assume $\lambda>0$, $p+q=M$, and $p,q\geq1$. Include all bidirectional incidence selectors, all $Q_{e\sigma}$, and cooling channels whose destination vectors span $\Hf$ for each spin. All included rates are strictly positive and time independent. If a Hermitian $H$ preserves the $(p,q)$ sector and satisfies $H|F_m\rangle=E_F|F_m\rangle$, then
\begin{equation}
 \Lcal_{\kappa,H}=-i[H,\cdot]+\Lcal_0+
 \sum_{e\sigma}\kappa_{e\sigma}\D[Q_{e\sigma}]
\label{eq:repaired}
\end{equation}
has $\PF$ as its unique stationary density matrix, and every initial density matrix in that sector converges to $\PF$ in trace norm.
\end{theorem}

The span condition is collective: for each spin, the full set of actual
destination vectors used by all sources must span $\Hf$; it does not require
each source separately to possess a complete destination set. Independent
endpoint cooling satisfies this condition. So does the four-cell periodic
single-exit protocol used below: with $e=j\mapsto v=j$, its aggregate
destinations are $\{f_0,f_1,f_2,f_3\}$, which span the four-dimensional
$\Hf$. Together with the degree-two periodic graph, all selectors and
dephasers, positive rates, and $H=0$, this protocol is an instance of the
theorem. For coherent destination combinations, one must check the span of
the actual vectors; counting channels is insufficient. The allowed
Hamiltonians include Eq.~\eqref{eq:parent} at arbitrary fixed strength. The
permanent obstruction proved above retains its separate assumption $H=0$.

The distinction between invariance and attraction is central to Markovian stabilization~\cite{Ticozzi2008Subsystems,Ticozzi2009Attractive}. Quasi-local stabilization and frustration-free fixed-point theory likewise require more than identifying a desired common kernel~\cite{Ticozzi2012Stabilizing,Johnson2016FixedPoints}. Our repair is a constructive choice of additional channels within that broader control problem~\cite{Ticozzi2010Constructive}. The proof excludes an arbitrary nonzero subspace $V\subset|F_m\rangle^\perp$ invariant under all jumps. With $g_{ef}=\{\bar d_e,\bar d_f^\dagger\}$ and $E_{ef}=\bar d_e^\dagger\bar d_f$, the commutators
\begin{align}
 [Q_e,Q_f]&=g_{ef}E_{ef}-g_{fe}E_{fe},\nonumber\\
 [Q_e,[Q_e,Q_f]]&=g_{ef}E_{ef}+g_{fe}E_{fe}-2|g_{ef}|^2Q_e
\end{align}
generate bright bilinears along overlapping edges. Connectivity extends them to all bright bilinears. Consequently, each bright-number block of $V$ contains the full bright Fock factor paired with a flat-coefficient space. Commuting cooling with that algebra replaces its source by any bright direction. The destination-span assumption then gives all bright-to-flat transfers as invariant-space constraints.

Choose a nonzero block with maximal total bright number. If a flat coefficient has a doublon contraction, a selector can create two additional bright particles. Minimum degree two ensures that its two projected creation directions are independent and that enough bright modes are available. If the maximal block has both flat and bright particles but no flat doublon, cooling can first create a nonzero doublon contraction; a selector then raises the total bright number above the supposed maximum. The all-bright block would generate the entire flat sector, including the target, and the all-flat block is reduced to the target by the selector constraints. Each possibility contradicts $V\subset|F_m\rangle^\perp$.

The same argument excludes mixed stationary states. If a stationary positive density matrix had a nonzero block orthogonal to the target, positivity of leakage would force that block's support to be invariant under all jumps. A Hamiltonian that preserves the target cannot cancel these nonnegative leakage terms. Target fidelity is monotone, and uniqueness of the stationary state forces its time average to approach one. Fidelity itself therefore tends to one. The trace-distance bound $D(\rho,\PF)\leq\sqrt{1-F}$~\cite{Fuchs1999Distance}, with the squared-fidelity convention $F=\Tr(\PF\rho)$, gives convergence in trace norm. We use the established invariant-support criterion~\cite{Kraus2008Preparation,Ticozzi2009Attractive,Ticozzi2014Quasilocal}; Supplemental Material Sec.~S6 proves the required algebra and support properties for this model.

Classical results on equilibrium and irreducibility of quantum semigroups provide useful algebraic context~\cite{Frigerio1977Equilibrium,Evans1977Irreducible}, but their rank and support hypotheses matter. In particular, Evans and Davies irreducibility must be distinguished~\cite{Zhang2024Davies}, and uniqueness criteria involving a faithful stationary state cannot be applied directly to our pure target~\cite{Nigro2019Uniqueness}. The absorbing target makes the full dynamics reducible in the Davies sense; the argument above instead rules out invariant supports orthogonal to it.

\section{Strong measurement and preparation times}
\label{sec:rates}

\subsection{The strong-measurement limit remains attractive}

Strong measurement can confine evolution to Zeno subspaces~\cite{Facchi2002Subspaces}, and strong dissipation can retain nontrivial slow dynamics within a steady-state manifold~\cite{Zanardi2014Manifolds,Popkov2018EffectiveZeno}. Write $\Lcal_\kappa=\kappa\Qcal+\Acal$, with $\Qcal=\sum_{e\sigma}r_{e\sigma}\D[Q_{e\sigma}]$ and fixed $r_{e\sigma}>0$. Keep the graph, Hamiltonian, and all other rates fixed as $\kappa\to\infty$. Since
\begin{equation}
 \langle X,\Qcal X\rangle_{\rm HS}
 =-\frac12\sum_{e\sigma}r_{e\sigma}\|[Q_{e\sigma},X]\|_{\rm HS}^2,
\end{equation}
the kernel is the commutant of all bright occupations. The algebra used above identifies its orthogonal projection as the conditional expectation
\begin{equation}
 \Ecal(X)=\bigoplus_{h_\uparrow,h_\downarrow}
 \frac{I_{b,h}}{d_{b,h}}\otimes\Tr_b(\Pi_hX\Pi_h),
\label{eq:expectation}
\end{equation}
where $d_{b,h}=\binom B{h_\uparrow}\binom B{h_\downarrow}$. The expectation depolarizes the bright factor within each bright-number block and retains the flat coefficients. Individual compact-orbital occupation patterns are not conserved by these overlapping measurements. The use of a fixed-point algebra and a projection onto it follows the operator-algebraic approach to stationary states~\cite{Frigerio1978Stationary}. Dissipation-projected dynamics~\cite{Zanardi2015Geometry} and structure-preserving adiabatic elimination~\cite{Azouit2016Elimination} provide related effective descriptions; the bright Fock factors in Eq.~\eqref{eq:expectation} are identified by this model's measurement algebra.

The finite-dimensional strong-dissipation limit~\cite{Burgarth2019StrongCoupling} gives
\begin{align}
 e^{t\Lcal_\kappa}&=e^{t\Lcal_Z}\Ecal+O(\kappa^{-1}),\nonumber\\
 \Lcal_Z&=\Ecal\Acal\Ecal\big|_{\operatorname{ran}\Ecal},
\label{eq:zeno}
\end{align}
uniformly on a fixed interval $0<t_0\leq t\leq T$, after the initial fast layer. Uniqueness of the compressed stationary state requires a further argument.

For this model, suppose that a stationary density matrix of $\Lcal_Z$ has a nonzero wrong support. It is invariant under all bright-mode unitaries because it lies in $\operatorname{ran}\Ecal$. Average $\Acal$ over that unitary group. The result is a valid Lindblad generator and agrees with $\Ecal\Acal\Ecal$ on invariant states. Positivity of target leakage and support leakage forces the wrong support to be invariant under every rotated jump. Continuity of the group action includes the original jumps. It is already invariant under all $Q_{e\sigma}$, so the recovery theorem excludes it. Thus $\PF$ remains globally attractive under $\Lcal_Z$ (Supplemental Material Sec.~S7).

Fast and slow spectral sectors also arise in analyses of dissipative Zeno limits~\cite{Popkov2021Spectrum}. Define a relaxing generator's gap by $g(\Lcal)=\min_{z\in\spec\Lcal\setminus\{0\}}[-\operatorname{Re}z]$. The slow eigenvalues approach the spectrum of $\Lcal_Z$, while the fast ones move to negative real parts of order $\kappa$. Therefore
\begin{equation}
 \lim_{\kappa\to\infty}g(\Lcal_\kappa)=g(\Lcal_Z)>0
 \qquad\text{at fixed }M.
\label{eq:gaplimit}
\end{equation}
The limit holds at fixed graph. A uniform thermodynamic gap and monotone acceleration with $\kappa$ are stronger claims that this argument does not establish. An $O(\kappa^{-1})$ error for individual eigenvalues would also require additional spectral regularity. A positive finite-size gap alone does not control size-dependent convergence prefactors or cutoff behavior~\cite{Szehr2015Spectral,Kastoryano2012Cutoff}; many-body examples also distinguish the gap from the relaxation time~\cite{Mori2020GapTime}.

\subsection{Finite-time upper and lower bounds}

The invariant-space proof supplies a finite-size convergence certificate without assuming a gap estimate. Absorb rates into jumps, fix a reference frequency $\Gamma$, and put $\widehat J_\mu=J_\mu/\sqrt\Gamma$. Reverse jump words of length at most $m\leq\mathscr D-1$ span the Hilbert space of dimension $\mathscr D=\binom{M+B}{p}\binom{M+B}{q}$. Hence
\begin{equation}
 G_m=\sum_{n=0}^m\sum_{|w|=n}w^\dagger\PF w>0,
\end{equation}
where $w$ is a product of the dimensionless jumps and the empty word is included. A quantum-jump expansion~\cite{Dalibard1992Trajectories,Dum1992MonteCarlo} yields computable $\tau>0$ and $\eta>0$ such that
\begin{equation}
 D(\rho(t),\PF)\leq\sqrt{1-F(0)}
 (1-\eta)^{\lfloor\Gamma t/\tau\rfloor/2}.
\label{eq:certificate}
\end{equation}
Supplemental Material Sec.~S8 gives the constants in terms of $\lambda_{\min}(G_m)$ and operator norms. These constants may deteriorate rapidly with dimension, so they do not give a polynomial bound on preparation time. Quantum $\chi^2$-divergence and logarithmic Sobolev methods provide other routes to mixing-time bounds~\cite{Temme2010ChiSquare,Kastoryano2013LogSobolev}, but their stationary-state support and full-rank assumptions require separate treatment for a pure absorbing target. We do not invoke such bounds here.

Weak dephasing gives a lower bound on the preparation time. For the chain state of Eq.~\eqref{eq:chain-initial}, take uniform rates $\kappa_{j\sigma}=\kappa$ and a weak parent Hamiltonian. The bright density then satisfies
\begin{multline}
 \frac{\langle N_b\rangle_t}{L}\geq
 \left(\frac12-\frac{w+4}{L}\right)\\
 \times\left[c_S-(w+3)\left(\frac\kappa2+4|\epsilon_H|\right)t\right]_+.
\label{eq:weak}
\end{multline}
Here $H=\epsilon_H H_{\rm MT}$, with the normalized chain convention $H_{\rm MT}=4\sum_{j\sigma}Q_{j\sigma}+U\sum_xq_x$. Only $O(w)$ boundary projectors have nonzero variance in the dark witness, which makes the drift bound independent of $L$. In particular, at $H=0$ reaching $F\geq1-\varepsilon_F$ with $\varepsilon_F<c_S$ requires
\begin{equation}
 t\geq\frac{2(c_S-\varepsilon_F)}{(w+3)\kappa}.
\end{equation}
This bounds the preparation time from the specified initial state without assuming a first-order law for the spectral gap.

Local particle transport imposes a separate limitation. Finite propagation-speed estimates originate in local Hamiltonian dynamics~\cite{Lieb1972Velocity} and extend to irreversible dynamics. Let $X_w=\sum_xw_xn_x$, with $w_x$ the distance to the initial occupied support. A bounded-degree, bounded-range number-conserving generator with fixed local strengths changes $\langle X_w\rangle$ at a rate of order $M$. The target must populate a macroscopic empty region, requiring a change of order $Mn$ on a graph of linear size $n$. At fixed sufficiently small target trace-distance error, $t_\varepsilon\geq c n=cM^{1/d}$. This elementary moment bound is consistent with locality bounds for Markovian dynamics~\cite{Poulin2010Locality,Barthel2012Quasilocality,Nachtergaele2011Irreversible}; it neither assumes diffusion nor determines an optimal dynamic exponent.

\section{Numerical verification}
\label{sec:numerics}

We test the static formulas on large graphs and the dynamics on small Fock spaces. All bright occupations use the orthogonal projector in Eq.~\eqref{eq:bright-number}. We evaluate static overlaps using Gram determinants and orthonormalized occupied subspaces, independently of the scalar boundary recurrences. The chain tests cover four offset sets, balanced and imbalanced sectors, and increasing lengths; decorated-square tests check the physical bright bound and area product. Figure~\ref{fig:static} compares these calculations with the analytic expressions. The static tests require only one-particle matrices. This use of Slater determinants is consistent with the tractability of fermionic linear optics~\cite{Bravyi2012FermionicOptics}, but does not make the full evolution Gaussian. Our four-fermion selectors, in particular, lie outside the quasi-free class, and bilinear cooling jumps do not by themselves imply covariance-matrix closure~\cite{Barthel2022Quadratic}. We therefore retain many-body Fock-space dynamics in the computations below.

For the smallest periodic graph used in the spectral comparison, $M=B=3$, $(p,q)=(2,1)$, and $\mathscr D=90$. We set $\lambda=1/\sqrt2$. The 12 endpoint cooling channels use $\bar f_v^\dagger\bar d_e$, each at rate one; the 12 directed selectors use Eq.~\eqref{eq:selector}, also at rate one. Each of the six $Q_{e\sigma}$ channels has rate $\kappa$. The actual Hamiltonian is $H=hH_{\rm par}$ with $t_0=1$, $U=1.7$, and the unnormalized $d_e$ of Eq.~\eqref{eq:orbitals}. Thus $h$ is not the coefficient of the differently normalized $H_{\rm MT}$ in Eq.~\eqref{eq:weak}.

\begin{figure*}[t]
\centering
\figfile{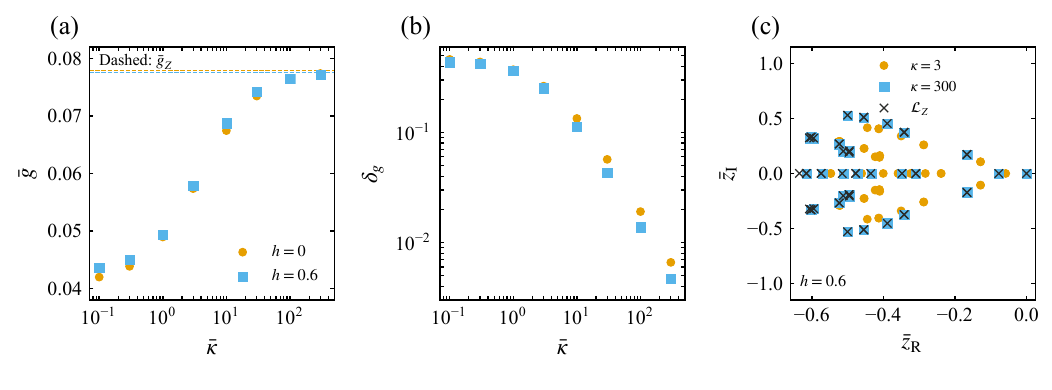}{17.8cm}
\caption{Fixed-size strong-dephasing comparison for the three-cell endpoint protocol at $\lambda=1/\sqrt2$, $p=2,q=1$. The axes use $\bar\kappa=\kappa/\gamma$, $\bar g=g/\gamma$, $\delta_g=|g-g_Z|/g_Z$, and $\bar z_{R,I}=(\operatorname{Re}z,\operatorname{Im}z)/\gamma$. (a) Full Liouvillian gaps, determined by the largest nonzero real part across all three translation sectors, for $h=0$ and $0.6$. Dashed lines are the gaps of the $190$-dimensional compression, $g_Z=0.0778747773$ and $0.0775354346$. (b) Relative difference $\delta_g$ from the corresponding $g_Z$. (c) A near-axis spectral window at $h=0.6$: the $16$ rightmost roots per translation sector at $\kappa=3$ and $300$ are compared with the compressed spectrum (crosses); degeneracies can overlap. The reference rate is the per-channel cooling rate $\gamma=1$, and $U=1.7$. Other parameters are fixed while $\kappa$ varies. The data test Eq.~\eqref{eq:gaplimit} at $M=3$, without establishing a thermodynamic gap, a universal error exponent, or monotonicity at arbitrary size.}
\label{fig:zeno}
\end{figure*}

The full Liouvillian has dimension $8100$. Translation acts as a symmetry of the superoperator by permuting equal-rate jumps; this is the weak-symmetry reduction of the Lindblad equation~\cite{Buca2012Symmetry,Albert2014Symmetries}. It decomposes the Liouvillian into three sectors of dimension $2700$, all of which are included in the rightmost-spectrum search. The compressed generator has operator dimension $190$. Its gaps are
\begin{align}
 g_Z(h=0)&=0.0778747773,\nonumber\\
 g_Z(h=0.6)&=0.0775354346.
\label{eq:gznumeric}
\end{align}
For $\kappa=300$, the full gaps are $0.0773602731$ and $0.0771691029$, respectively. Figure~\ref{fig:zeno} shows their approach to Eq.~\eqref{eq:gznumeric}. The finite-rate calculations find one zero root across the symmetry sectors, with eigenpair residuals below $8\times10^{-12}$. These floating-point residuals check the computed eigenpairs; they do not certify the full spectrum by interval arithmetic.

\begin{figure*}[t]
\centering
\figfile{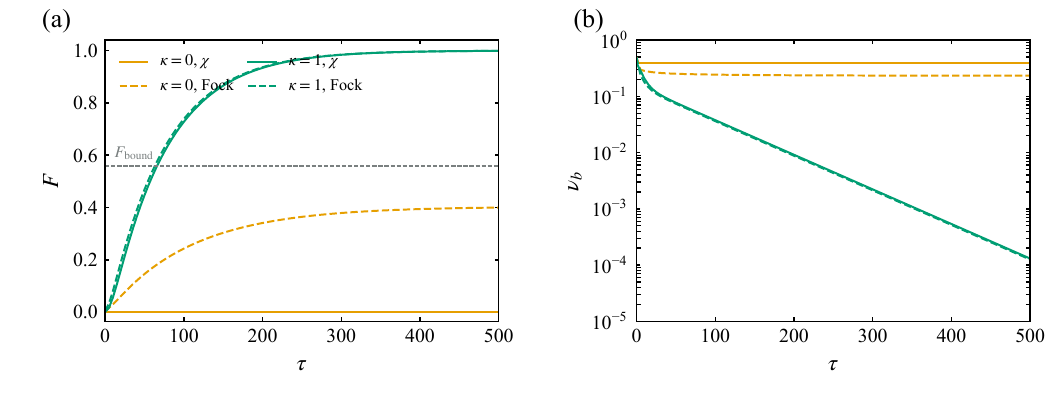}{17.8cm}
\caption{Long-time four-cell single-exit dynamics at $p=q=2$ and $H=0$ through $\tau=\gamma t=500$. (a) Target fidelity $F$; (b) normalized physical bright population $\nu_b=\langle N_b\rangle/M$. Solid curves start from the target-orthogonal wrong dark state $\chi$, and dashed curves start from the physical Fock state with up spins on $A_0,B_0$ and down spins on $A_2,B_2$. Orange and green denote $\kappa=0$ and $1$, respectively. The gray dashed line in (a) is the all-time bound $F\leq0.5589846446$ for the Fock state at $\kappa=0$ only. The long trajectories use the exact translation-twirl reduction: the reconstructed density matrix is the translation-averaged state, while the displayed translation-invariant observables agree with unreduced evolution at shared times through $\tau=100$ within $1.776\times10^{-15}$. At $\kappa=1$, $F_\chi(500)=0.9990602765$, $\nu_b(500)=1.303585010\times10^{-4}$, and $\|\mathbb L\operatorname{vec}\rho(500)\|_2=1.3527589152187228\times10^{-5}$. These are finite-time data and quality checks supporting the recovery mechanism, not a numerical plateau or a proof of attraction; global attraction follows analytically from Theorem~2. This small graph is outside the $a,b\geq3$ macroscopic sequence. Each cooling and selector channel has rate $\gamma=\eta=1$.}
\label{fig:dynamics}
\end{figure*}

The three-vertex endpoint graph already has only one common pure dark state at $\kappa=0$, so its time traces cannot demonstrate removal of a wrong pure dark state. A distinct four-vertex single-exit chain provides that test. In the $(2,2)$ sector, start from $f_{0\uparrow}^\dagger u_{0\uparrow}^\dagger f_{2\downarrow}^\dagger u_{2\downarrow}^\dagger|0\rangle$, project out its target component, and normalize. Its cooling and selector residuals are below $5\times10^{-17}$, its target overlap is below $6\times10^{-17}$, and its initial bright density is $0.3913043478$. This finite-size witness is exactly stationary without dephasing. Figure~\ref{fig:dynamics} compares it with the repaired dynamics. Extending the $\kappa=1$ trajectory to $t=500$ gives $F_\chi=0.9990602765$ and $\langle N_b\rangle/M=0.0001303585$ (Supplemental Material Sec.~S9). The endpoint residual $\|\mathbb L\operatorname{vec}\rho(500)\|_2=1.3527589152187228\times10^{-5}$ makes this a finite-time quality check, not a stationary numerical plateau. The numerical calculations record trace, Hermiticity, positivity, and target residual checks.

For the three-vertex graph, the reverse-reachable dimensions are
$1,22,89,90$, and the word Gram matrix first reaches full rank at $m=3$.
Evaluating Eq.~\eqref{eq:certificate} gives strictly positive finite-size
constants, but the resulting guaranteed time is many orders of magnitude
longer than the observed relaxation time. Supplemental Material Sec.~S9
reports the floating-point values and their numerical-status caveat. Thus
the jump-word construction is an existence certificate, not an effective
mixing-time estimate.

\section{Discussion}

The obstruction arises because the number of source particles scales with volume while the available residual flat directions are limited by a boundary rank. This converts the geometry of nonorthogonal compact orbitals into a bound on the physical bright number. Combining that bound with an explicit initial overlap constrains the entire preparation dynamics. In higher dimensions, the overlap depends strongly on boundary preparation: the bare physical Fock state and the state after local boundary rotations can have parametrically different weights in the same dark witness.

Bright-occupation dephasing removes the wrong invariant subspaces because the measured modes overlap. Their noncommuting projectors generate rearrangements that make blocked transfers accessible. Each channel acts within a compact orbital's support. On a decorated chain, implementing it requires a coupling to a coherent three-site mode; measuring the density of a single physical site implements a different channel. The theorem specifies the required mode measurements, while their realization on a particular platform remains to be developed. Optical Lieb-lattice experiments demonstrate coherent flat-band dynamics and interaction-induced band distortions~\cite{Taie2015OpticalLieb,Ozawa2017Interaction}, but those bosonic settings do not implement the present fermionic channels. Dissipation-generated entanglement in atomic ensembles provides a further experimental precedent~\cite{Krauter2011Entanglement}; its use of measurement records for prolonged stabilization should be distinguished from our unread measurements. Effective-operator elimination of lossy auxiliary levels~\cite{Reiter2012Effective} and repeated-ancilla collision models~\cite{Cattaneo2021Collision} are possible design tools. A microscopic construction respecting this protocol's compact supports, conservation laws, and rates remains necessary.

In the strong-dephasing limit, direct dissipative transfers survive the conditional expectation while the bright factor is mixed at fixed bright number. The projected generator has no wrong absorbing support, so the slow gap remains positive at fixed graph. This differs from suppression of coherent transitions in the quantum Zeno effect~\cite{Misra1977Zeno,Facchi2002Subspaces}. The gap may nevertheless close with graph size, and the jump-word certificate is too conservative to determine its scaling. Studies of algebraic versus exponential decoherence~\cite{Cai2013Decoherence} and exact spectra of chains with site-density dephasing~\cite{Medvedyeva2016Dephasing} illustrate the dependence on the model and channels. Neither their asymptotic laws nor their dephasing operators can be transferred to the overlapping compact-mode measurements used here. For local implementations and the constructed inhomogeneous initial states, the finite-size upper bound and the transport lower bound leave a wide range of possible preparation times.

Other flat bands require their own Gram and support analysis. Kagome Hubbard models~\cite{Mielke1992Kagome} and chiral flat-band constructions~\cite{Ramachandran2017Chiral} offer possible geometries, but need not share our independent orbital frame or selector structure. Band touching and singular Bloch eigenvectors can affect the completeness of compact localized states~\cite{Bergman2008BandTouching,Rhim2019Classification,Rhim2021Singular}. These issues do not invalidate the present frame, whose unit block guarantees independence, but must be reconsidered before extending the result.

Dissipative topological wires~\cite{Diehl2011Topology}, more general topological steady states~\cite{Bardyn2013Topology}, and Chern-insulator preparation~\cite{Budich2015Chern} raise related questions about attainable targets. The no-go theorem for finite-range quadratic Lindblad constructions constrains simultaneous nonzero Chern topology, exact purity, and a finite relaxation rate under its stated hypotheses~\cite{Goldstein2019NoGo}. Non-equilibrium topological response is also studied through field theory~\cite{Tonielli2020FieldTheory}. Our obstruction uses Pauli blocking and a compact-orbital Gram cut, and neither assumes a Chern invariant nor establishes a general topological preparation bound.

The graph and channel assumptions delimit both conclusions. The permanent obstruction assumes $H=0$, whereas recovery permits any $H$ that preserves the target and particle sector. Equilibrium stability of Hubbard ferromagnetism beyond ideal flat bands~\cite{Tasaki1995Hubbard,Tasaki1996Stability} does not establish persistence of our exact wrong dark states under dispersive perturbations. The sufficient recovery criterion established here assumes all specified bright dephasers, a collectively spanning set of actual cooling destinations for each spin, and the stated graph degree; we do not claim that these conditions are necessary or minimal. Removing channels or allowing a finite density of preparation errors remains an open extension. Stability results for local dissipative systems suggest a framework~\cite{Cubitt2015Stability}, but their mixing assumptions have not been established for this graph family. The rank bound quantifies the blocked bright population, and the measurement algebra identifies channels that restore global attraction.

\section*{Data and code availability}
The Supplemental Material provides the analytic derivations, numerical
methods, and validation results. The numerical datasets, analysis scripts,
and validation records are retained in a separate research archive.

\bibliography{references}
\end{document}


\maketitle
\vspace{-2em}
\begin{abstract}
This supplement gives self-contained proofs for the independent-transfer kernel, the boundary-rank spectral bound, explicit chain and boundary-prepared initial states, measurement-assisted global attraction, the fixed-size strong-dephasing limit, and finite-time bounds. It fixes the distinction between nonorthogonal orbital coordinates and physical particle numbers, and between the normalized chain and unnormalized graph Hamiltonians. The final section describes computational conventions and validation checks.
\end{abstract}
\begingroup\footnotesize
\tableofcontents
\endgroup
\clearpage

\section{Physical orbitals, selector kernel, and the ferromagnetic target}
\label{sup:geometry}

Let $G=(V,E)$ be a finite connected simple graph, with $M=|V|$ vertices and
$B=|E|$ edges. Its decorated physical graph has a site $A_v$ for each vertex
and a site $B_e$ for each edge. Each physical site carries two fermionic spin
modes. We work in the sector
\begin{equation}
 N_\uparrow=p,\qquad N_\downarrow=q,\qquad p+q=M,\qquad p,q\geq1.
 \label{sup:sector}
\end{equation}
Thus the spin-degenerate flat space is half filled. This is not half filling
of all $M+B$ physical sites. All scalar products below are physical Fock-space
scalar products; coefficients in a nonorthogonal orbital basis are never
interpreted directly as probabilities.

For $\lambda>0$ and $e=(v,w)$, the unnormalized orbital creation operators are
\begin{align}
 f_{v\sigma}^\dagger
 &=c_{A_v,\sigma}^\dagger-\lambda\sum_{e\ni v}c_{B_e,\sigma}^\dagger,
 &u_{e\sigma}^\dagger&=c_{B_e,\sigma}^\dagger,\\
 d_{e\sigma}^\dagger
 &=c_{B_e,\sigma}^\dagger
   +\lambda(c_{A_v,\sigma}^\dagger+c_{A_w,\sigma}^\dagger).
 \label{sup:orbitals}
\end{align}
If $\mathsf B$ is the unsigned vertex-edge incidence matrix, the corresponding
column frames and their Gram matrices are
\begin{align}
 \mathsf F&=\begin{pmatrix}I_M\\-\lambda\mathsf B^{\mathsf T}\end{pmatrix},
 &\mathsf D&=\begin{pmatrix}\lambda\mathsf B\\I_B\end{pmatrix},\\
 G_f&=I_M+\lambda^2\mathsf B\mathsf B^{\mathsf T},
 &G_b&=I_B+\lambda^2\mathsf B^{\mathsf T}\mathsf B.
 \label{sup:grams}
\end{align}
Both Gram matrices are positive definite, while
$\mathsf F^\dagger\mathsf D=0$. Consequently the flat and bright spaces
$H_f=\operatorname{ran}\mathsf F$ and $H_b=\operatorname{ran}\mathsf D$ are
orthogonal complements, of dimensions $M$ and $B$. Their orthogonal
projectors are
\begin{equation}
 \Pi_f=\mathsf F G_f^{-1}\mathsf F^\dagger,\qquad
 \Pi_b=\mathsf D G_b^{-1}\mathsf D^\dagger=I-\Pi_f.
 \label{sup:projectors}
\end{equation}
The observable used throughout is the orthogonal bright number
\begin{equation}
 N_b=\sum_\sigma c_\sigma^\dagger\Pi_b c_\sigma.
 \label{sup:bright-number}
\end{equation}
The number of $u_e$ labels in an exterior-basis monomial is generally different
from its bright population.

The selector on an oriented incidence edge $x\to y$ is
\begin{equation}
 P_{x\to y}=T_{xy}^\dagger b_x,\qquad
 b_x=c_{x\downarrow}c_{x\uparrow},\qquad
 T_{xy}^\dagger=\frac{c_{x\uparrow}^\dagger c_{y\downarrow}^\dagger+
 c_{x\downarrow}^\dagger c_{y\uparrow}^\dagger}{\sqrt2}.
 \label{sup:selector}
\end{equation}
Both orientations of every incidence edge joining $A_v$ to $B_e$ are included. Additional
selectors of the same form may be added if needed; they annihilate every
state without physical doublons.

\begin{lemma}[Selector kernel]
\label{sup:selector-kernel}
In the sector \eqref{sup:sector}, the common kernel of the bidirectional
incidence selectors is exactly $\bigcap_x\ker b_x$.
\end{lemma}
\begin{proof}
Write $q_x=b_x^\dagger b_x=n_{x\uparrow}n_{x\downarrow}$. After $b_x$ acts,
site $x$ is empty. Direct fermionic contraction on this subspace gives
\begin{equation}
 T_{xy}T_{xy}^\dagger
 =1-\frac{n_{y\uparrow}+n_{y\downarrow}}2,
 \qquad
 \|P_{x\to y}\psi\|^2
 =\left\langle q_x\left(1-\frac{n_{y\uparrow}+n_{y\downarrow}}2\right)
 \right\rangle_\psi.
 \label{sup:selector-norm}
\end{equation}
The diagonal operator inside the expectation is nonnegative. Its vanishing
means that every Fock configuration in $\psi$ that is doubly occupied at $x$
is also doubly occupied at $y$. In projection notation,
$q_x(1-q_y)\psi=0$. The reverse selector gives
$q_y(1-q_x)\psi=0$, and hence $q_x\psi=q_y\psi$. Connectivity propagates this
identity to all physical sites. A common nonzero doubly occupied component
would have every physical site doubly occupied, requiring $2(M+B)$ particles.
That component is absent when $N=M$. Thus $b_x\psi=0$ for every $x$.
The converse follows immediately from \eqref{sup:selector}.
\end{proof}

\begin{lemma}[Unique target in the flat space]
\label{sup:flat-target}
The strictly flat states in \eqref{sup:sector} with no physical doublons form
a one-dimensional space. Define the physical global lowering operator by
$S^-=\sum_{v\in V}c_{A_v\downarrow}^\dagger c_{A_v\uparrow}
+\sum_{e\in E}c_{B_e\downarrow}^\dagger c_{B_e\uparrow}$, where the sums run
over every physical vertex and edge site. Its normalized vector is
\begin{equation}
 |F_m\rangle=\sqrt{\frac{p!}{q!M!}}(S^-)^q|\Omega_\uparrow\rangle,
 \qquad
 |\Omega_\uparrow\rangle=\frac{\prod_{v\in V}f_{v\uparrow}^\dagger|0\rangle}
 {\sqrt{\det G_f}},\qquad m=(p-q)/2.
 \label{sup:target}
\end{equation}
\end{lemma}
\begin{proof}
Expand a flat state in the exterior basis of the independent $f_v$ orbitals.
The physical annihilator at $A_v$ contracts only the label $v$. The constraints
$b_{A_v}\psi=0$ therefore exclude simultaneous up and down occupation of any
one flat label. There are $M$ particles and $M$ available labels, so each label
is occupied once. The remaining coefficients can be indexed by spin words of
length $M$ with $p$ up spins and $q$ down spins, using one fixed order of labels.

For $e=(v,w)$, the operator at $B_e$ contracts only the flat labels $v,w$, each
with coefficient $-\lambda$. The constraint $b_{B_e}\psi=0$ equates the
coefficients of the two words obtained by interchanging opposite spins at
$v,w$. The relative minus sign from fermionic reordering converts the
doublon-cancellation equation into equality of these coefficients. Edge
transpositions of a connected graph generate all permutations of its
vertices. Every spin word with the specified magnetization therefore has the
same coefficient. This is the maximal-spin state in \eqref{sup:target}.
The normalization follows from the usual spin-$M/2$ lowering matrix element,
$\|(S^-)^q|\Omega_\uparrow\rangle\|^2=q!M!/p!$.
This is the connectivity argument for flat-band ferromagnetism in the present
orbital coordinates \cite{Tasaki1998Review}.
\end{proof}

We write $\PF=|F_m\rangle\langle F_m|$ and
$F(t)=\Tr[\PF\rho(t)]$. The positive parent Hamiltonian
\begin{equation}
 H_{\rm par}=t_0\sum_{e\sigma}d_{e\sigma}^\dagger d_{e\sigma}
       +U\sum_xq_x,\qquad t_0,U>0,
 \label{sup:parent}
\end{equation}
annihilates the target. Moreover, $G_b\geq I_B$ implies
$H_{\rm par}\geq t_0N_b$. Until Section~\ref{sup:recovery}, this Hamiltonian
is only an energy diagnostic: the actual Hamiltonian in every permanent
obstruction statement is zero.

\section{The complete kernel of independent transfer channels}
\label{sup:cooling-kernel}

Assign a destination set $\operatorname{Dest}(e)\subseteq V$ to each source
edge, allowing an empty set, and include each transfer
\begin{equation}
 C_{ev\sigma}=f_{v\sigma}^\dagger d_{e\sigma},\qquad
 v\in\operatorname{Dest}(e),
 \label{sup:independent-cooling}
\end{equation}
as a separate jump. Since $\{d_e,f_v^\dagger\}=0$ and
$\{d_e,u_{e'}^\dagger\}=\delta_{ee'}$, the $f_v,u_e$ orbitals are linearly
independent and together span the entire one-particle space. For one spin,
the vectors
\begin{equation}
 |I,J\rangle=f_I^\dagger u_J^\dagger|0\rangle,\qquad
 I\subseteq V,\quad J\subseteq E,\quad |I|+|J|=n,
 \label{sup:exterior-basis}
\end{equation}
are a nonorthogonal basis of the $n$-particle sector. Products throughout use
one fixed ordering convention.

\begin{theorem}[Independent-transfer kernel]
\label{sup:kernel-theorem}
Let $D(J)=\bigcup_{e\in J}\operatorname{Dest}(e)$. The common kernel of all
one-spin transfers \eqref{sup:independent-cooling} is
\begin{equation}
 \mathcal K_n=\operatorname{span}\{
 |I,J\rangle: |I|+|J|=n,\ D(J)\subseteq I\},
 \label{sup:exact-kernel}
\end{equation}
and its dimension is
\begin{equation}
 \dim\mathcal K_n=\sum_{J\subseteq E}
 \binom{M-|D(J)|}{n-|J|-|D(J)|},
 \label{sup:kernel-count}
\end{equation}
with out-of-range binomial coefficients understood as zero.
\end{theorem}
\begin{proof}
Fix one channel $(e,v)$. It acts nontrivially on \eqref{sup:exterior-basis}
precisely when $e\in J$ and $v\notin I$. Up to the ordering sign, its output is
$|I\cup\{v\},J\setminus\{e\}\rangle$. For this fixed channel, a nonzero output
has exactly one possible input: remove $v$ from the flat-label set and insert
$e$ into the source-label set. Outputs from distinct nonzero monomial inputs
therefore cannot cancel. Every coefficient violating $D(J)\subseteq I$ must
vanish in a common-kernel vector. Conversely, that inclusion makes every
channel zero by either the absence of its source or Pauli occupation of its
destination. For fixed $J$, all $|D(J)|$ required labels are occupied and the
remaining $n-|J|-|D(J)|$ flat labels can be selected freely. This proves
\eqref{sup:kernel-count}.
\end{proof}

The two-spin cooling kernel is $\mathcal K_p\otimes\mathcal K_q$, after
fixing the usual fermionic ordering convention between spin sectors. The
kernel of cooling and selectors is its intersection with the no-doublon
space. Equation~\eqref{sup:kernel-count} does not give the dimension of that
intersection. If several destinations are combined into a single coherent
jump, cancellation between destinations can enlarge the kernel, so
\eqref{sup:exact-kernel} need not remain an equality. A vector dark for every
individual destination remains dark for any such coherent combination. The
obstruction constructions below use only this latter, sufficient property.

\section{Chain Fock initial states and all-time preparation bounds}
\label{sup:chain}

\subsection{Normalization and the integer sequence}

For a cycle of length $L$, identify edge $j$ with $(j,j+1)$ and take indices
modulo $L$. In this section only, use normalized chain orbitals
\begin{align}
 f_{j\sigma}^\dagger&=\frac12c_{B_{j-1},\sigma}^\dagger
 -\frac1{\sqrt2}c_{A_j,\sigma}^\dagger
 +\frac12c_{B_j,\sigma}^\dagger,\\
 d_{j\sigma}^\dagger&=\frac12c_{A_j,\sigma}^\dagger
 +\frac1{\sqrt2}c_{B_j,\sigma}^\dagger
 +\frac12c_{A_{j+1},\sigma}^\dagger,
 &u_{j\sigma}^\dagger&=\sqrt2c_{B_j,\sigma}^\dagger.
 \label{sup:chain-orbitals}
\end{align}
At $\lambda=1/\sqrt2$, these are related to \eqref{sup:orbitals} by
$f_{\rm chain}=-f_{\rm graph}/\sqrt2$,
$d_{\rm chain}=d_{\rm graph}/\sqrt2$, and
$u_{\rm chain}=\sqrt2u_{\rm graph}$. The physical flat and bright projectors
are unchanged, while jump rates and Hamiltonian coefficients must be
converted if the two conventions are compared. The flat Gram matrix is
\begin{equation}
 (G_f)_{ij}=\delta_{ij}
       +\frac14(\delta_{i,j+1}+\delta_{i,j-1}).
 \label{sup:chain-gram}
\end{equation}

Fix a nonempty finite set $S\subset\mathbb Z$ of transfer offsets. It remains
fixed as $L$ grows. Define
\begin{equation}
 \ell_-=-\min(S\cup\{0\}),\qquad
 \ell_+=\max(S\cup\{0\}),\qquad w=\ell_-+\ell_+.
 \label{sup:footprint}
\end{equation}
The transfers are $C_{jr\sigma}=f_{j+r,\sigma}^\dagger d_{j\sigma}$,
$r\in S$. Choose integers $a,b\geq3$ and set
\begin{equation}
 p=2a+w,\qquad q=2b+w,\qquad L=2(a+b+w).
 \label{sup:chain-sequence}
\end{equation}
With $m_\uparrow=a+w$ and $m_\downarrow=b+w$, take the flat-label intervals
\begin{equation}
 I_\uparrow=[1,m_\uparrow],\qquad
 I_\downarrow=[m_\uparrow+2,m_\uparrow+m_\downarrow+1].
 \label{sup:chain-intervals}
\end{equation}
For $I_\sigma=[A_\sigma,B_\sigma]$, set
\begin{equation}
 J_\sigma=[A_\sigma+\ell_-,B_\sigma-\ell_+].
 \label{sup:chain-sources}
\end{equation}
The source sets have sizes $a,b$. The normalized physical Fock initial state
and a normalized dark witness are
\begin{align}
 |\Phi_S\rangle&=\prod_\sigma
 \left(\prod_{i\in I_\sigma}c_{A_i,\sigma}^\dagger\right)
 \left(\prod_{j\in J_\sigma}c_{B_j,\sigma}^\dagger\right)|0\rangle,
 \label{sup:chain-fock}\\
 |\Psi_S\rangle&=\frac1{\mathcal N_S}\prod_\sigma
 f_{I_\sigma,\sigma}^\dagger u_{J_\sigma,\sigma}^\dagger|0\rangle.
 \label{sup:chain-witness}
\end{align}
The constant $\mathcal N_S$ is the physical norm of the numerator. The
independence of the $f,u$ coordinates ensures that it is nonzero. Equation
\eqref{sup:chain-fock} specifies occupations of physical sites and therefore
requires no orbital superposition in the initial preparation.

\begin{theorem}[All-time chain obstruction]
\label{sup:chain-theorem}
Define $D_0=1$, $D_1=3/2$, and
\begin{equation}
 D_n=2D_{n-1}-\frac14D_{n-2}\quad(n\geq2),\qquad
 c_S=[D_{\ell_-+1}D_{\ell_+}]^{-2}.
 \label{sup:chain-constant}
\end{equation}
For the $H=0$ transfer-and-selector dynamics initialized in
\eqref{sup:chain-fock}, every $t\geq0$ obeys
\begin{equation}
 F(t)\leq1-c_S,\qquad
 \frac{\Tr[N_b\rho(t)]}{L}
 \geq c_S\left(\frac12-\frac{w+4}{L}\right).
 \label{sup:chain-bounds}
\end{equation}
The same bounds hold for locally integrable nonnegative time-dependent
rates and for coherent combinations of the individual dark channels.
\end{theorem}

\subsection{Darkness and the exact boundary determinant}

Every source $j\in J_\sigma$ has all destinations $j+S$ in $I_\sigma$. Thus
each transfer annihilates \eqref{sup:chain-witness}. The physical support of
the flat interval $[A,B]$ consists of $A_A,\ldots,A_B$ and
$B_{A-1},\ldots,B_B$. The two supports in \eqref{sup:chain-intervals} are
disjoint, including across the periodic boundary: the final down-spin label
is at most $L-1$ because $a,b\geq3$. No physical site carries both spins, and
every selector is zero.

For one spin, first wedge all occupied $B_j$ with $j\in J$. Their components
can then be deleted from each $f_i$ without changing the exterior product.
The remaining one-particle columns are
\begin{equation}
 g_i=-\frac{|A_i\rangle}{\sqrt2}
 +\frac12\sum_{e\in\{i-1,i\}\setminus J}|B_e\rangle,
 \qquad i\in I.
 \label{sup:residual-chain-columns}
\end{equation}
Let $C_{ie}=1$ when the unfilled edge $e$ equals $i-1$ or $i$, and zero
otherwise. The Gram matrix of these columns is
$H=I_{|I|}/2+CC^{\mathsf T}/4$. The probability of their component with all
the $A_i$ occupied is therefore
\begin{equation}
 \frac{(1/2)^{|I|}}{\det H}
 =\det\left(I+\frac12C^{\mathsf T}C\right)^{-1}.
 \label{sup:boundary-determinant}
\end{equation}
The equality is the determinant identity $\det(I+XY)=\det(I+YX)$.
The unfilled edge set has two separated blocks, of lengths
$\ell_-+1$ and $\ell_+$. Each block is tridiagonal, with off-diagonal entries
$1/2$, diagonal entries $2$ except for the outermost entry $3/2$.
Expanding its determinant at the inner end gives \eqref{sup:chain-constant}.
Each spin contributes $[D_{\ell_-+1}D_{\ell_+}]^{-1}$, so
\begin{equation}
 |\langle\Psi_S|\Phi_S\rangle|^2=c_S.
 \label{sup:chain-overlap}
\end{equation}
The determinant depends on the fixed boundary thickness, not the lengths of
the occupied interiors. For example, $c_{\{0\}}=4/9$,
$c_{\{0,1\}}=16/81$, and $c_{\{-1,0,1\}}=64/1089$.

\subsection{Bright spectral support and witness monotonicity}

Let $F_I=\operatorname{span}\{f_i:i\in I\}$, with physical orthogonal
projector $P_I$, and let
$R_I=(I-P_I)\operatorname{span}\{u_j:j\in J\}$. The dimension of $R_I$ is
$|J|$, because the $f,u$ coordinates are independent. For
$x=\sum_k t_k f_k\in H_f\cap F_I^\perp$, the orthogonality equations are
\begin{equation}
 t_{i-1}+4t_i+t_{i+1}=0,\qquad i\in I.
 \label{sup:chain-recurrence}
\end{equation}
Two boundary numbers determine the sequence on the interval and its adjacent
endpoints. Since
$\langle x,u_j\rangle=(t_j^*+t_{j+1}^*)/\sqrt2$, the map from $x$ to all
these source overlaps has rank at most two. This remains true if the source
$j=B$ is included when $\ell_+=0$: the recurrence at $B$ also fixes
$t_{B+1}$. Taking the adjoint of the overlap map yields
\begin{equation}
 \rank(\Pi_f|_{R_I})\leq2.
 \label{sup:chain-rank}
\end{equation}

Choose an orthogonal decomposition $R_I=K\oplus E$ with $K=R_I\cap H_b$.
Then $\dim E\leq2$. An exterior product of $a$ vectors in $R_I$ contains at
most two factors from $E$ and at least $a-2$ factors from $K$. The latter are
strictly bright. This argument also applies to arbitrary linear combinations
of such products. Filling the already occupied flat directions cannot change
their bright count. Applied to both spins, it gives
\begin{equation}
 |\Psi_S\rangle\in\operatorname{ran}\mathbf1_{[h,\infty)}(N_b),
 \qquad h=a+b-4=\frac L2-w-4>0.
 \label{sup:chain-spectral}
\end{equation}
Thus $\Psi_S$ is orthogonal to the target. With
$P_\Psi=|\Psi_S\rangle\langle\Psi_S|$, the operator inequalities are
\begin{equation}
 \PF+P_\Psi\leq I,\qquad
 N_b\geq h\mathbf1_{[h,\infty)}(N_b)\geq hP_\Psi.
 \label{sup:witness-operator}
\end{equation}
They do not require $P_\Psi$ to commute with $N_b$.

Absorb rates into the actual jumps $J_\mu$. Because $J_\mu P_\Psi=0$ and the
actual Hamiltonian is zero,
\begin{equation}
 \Lcal_0^\dagger(P_\Psi)
 =\sum_\mu J_\mu^\dagger P_\Psi J_\mu\geq0.
 \label{sup:witness-drift}
\end{equation}
The witness probability can only increase. Its initial value is $c_S$ by
\eqref{sup:chain-overlap}; combining this fact with
\eqref{sup:witness-operator} proves Theorem~\ref{sup:chain-theorem}.
The argument is pointwise in time and therefore also covers the stated rate
variations. Along \eqref{sup:chain-sequence}, with both spin densities bounded
away from zero, it follows that
\begin{equation}
 \liminf_{L\to\infty}\inf_{t\geq0}
 \frac{\Tr[N_b\rho(t)]}{L}\geq\frac{c_S}{2}>0.
 \label{sup:chain-thermo}
\end{equation}
These are lower bounds, not asserted values of a dynamical plateau. The
constant can decrease as the footprint grows; the theorem keeps $S$ fixed.

In the chain normalization, $B_d=\sum_{j\sigma}d_{j\sigma}^\dagger
d_{j\sigma}\geq N_b/2$. Hence the diagnostic Hamiltonian
$H_{\rm MT}=4B_d+U\sum_xq_x$ satisfies $H_{\rm MT}\geq2N_b$. Equation
\eqref{sup:chain-thermo} gives asymptotic density lower bounds $c_S/4$ for
$B_d$ and $c_S$ for $H_{\rm MT}$. If the initial density matrix differs from
$|\Phi_S\rangle\langle\Phi_S|$ by trace distance at most $\delta$, the same
reasoning replaces $c_S$ by $(c_S-\delta)_+$. This assumes control of the
global trace distance, not a fixed independent error at every site.

\section{Cut Gram rank and extensive dark subspaces on decorated graphs}
\label{sup:cut-rank}

Return to the graph convention \eqref{sup:orbitals}. Choose disjoint vertex
regions $U_\uparrow,U_\downarrow$ with no base-graph edge joining them, and set
\begin{equation}
 T(U)=\{e=(v,w):v,w\in U,\ \operatorname{Dest}(e)\subseteq U\}.
 \label{sup:legal-sources}
\end{equation}
Select integers $a_\sigma$ such that
\begin{equation}
 0\leq a_\sigma\leq|T(U_\sigma)|,\qquad
 |U_\uparrow|+a_\uparrow=p,\qquad
 |U_\downarrow|+a_\downarrow=q.
 \label{sup:region-filling}
\end{equation}
Let $W$ be the span of the exterior-basis vectors
\begin{equation}
 \prod_\sigma f_{U_\sigma,\sigma}^\dagger
 u_{J_\sigma,\sigma}^\dagger|0\rangle,
 \qquad J_\sigma\subseteq T(U_\sigma),\quad|J_\sigma|=a_\sigma.
 \label{sup:graph-witness-space}
\end{equation}
Transfers vanish destination by destination, and the two spin supports are
disjoint, so all selectors also vanish. The independence of the exterior
basis gives the exact dimension of this constructed subspace:
\begin{equation}
 \dim W=\prod_{\sigma=\uparrow,\downarrow}
 \binom{|T(U_\sigma)|}{a_\sigma}.
 \label{sup:W-dimension}
\end{equation}

\begin{lemma}[Cut-rank bound]
\label{sup:rank-lemma}
Let $F_U=\operatorname{span}\{f_v:v\in U\}$, let $P_U$ be its physical
orthogonal projector, and define
\begin{equation}
 R_U=(I-P_U)\operatorname{span}\{u_e:e\in T(U)\},\qquad
 r(U)=\rank(G_f)_{U,U^c}.
 \label{sup:cut-definition}
\end{equation}
Then $\rank(\Pi_f|_{R_U})\leq r(U)\leq|\partial U|$, where $\partial U$
is the set of vertices in $U$ incident on a cut edge.
\end{lemma}
\begin{proof}
For $x=\mathsf F c\in H_f\cap F_U^\perp$, orthogonality gives
\begin{equation}
 (G_f)_{U,U}c_U+(G_f)_{U,U^c}c_{U^c}=0,
 \qquad
 c_U=-(G_f)_{U,U}^{-1}(G_f)_{U,U^c}c_{U^c}.
 \label{sup:cut-equation}
\end{equation}
The principal Gram block is positive definite and invertible. Thus $c_U$
depends on at most $r(U)$ independent combinations of exterior coefficients.
For an internal source edge $e=(v,w)$,
$\langle x,u_e\rangle=-\lambda(c_v^*+c_w^*)$ reads only $c_U$.
The map from $H_f\cap F_U^\perp$ to the residual-source overlaps consequently
has rank at most $r(U)$. Its adjoint is the flat projection of $R_U$ into
$H_f\cap F_U^\perp$, which proves the first inequality. Rows of
$(G_f)_{U,U^c}$ away from $\partial U$ vanish, proving the second.
\end{proof}

\begin{theorem}[Bright spectral support of the dark subspace]
\label{sup:graph-theorem}
The whole subspace \eqref{sup:graph-witness-space} satisfies
\begin{equation}
 W\subseteq\operatorname{ran}\mathbf1_{[h_G,a_\uparrow+a_\downarrow]}(N_b),
 \qquad h_G=\sum_\sigma\bigl(a_\sigma-r(U_\sigma)\bigr)_+.
 \label{sup:graph-spectral}
\end{equation}
If $h_G>0$, it is orthogonal to the target, and every density matrix supported
on $W$ is stationary for the original $H=0$ protocol.
\end{theorem}
\begin{proof}
Subtracting $P_Uu_e$ from a source orbital leaves the exterior product with
all $f_U$ unchanged. The residual-source map is injective by independence of
the $f,u$ basis. Decompose $R_U=K\oplus E$, where $K=R_U\cap H_b$ and $E$ is
its orthogonal complement in $R_U$. Lemma~\ref{sup:rank-lemma} gives
$\dim E\leq r(U)$. Any term in $\bigwedge^aR_U$ has at most $r(U)$ factors
from $E$, so it has at least $(a-r(U))_+$ strictly bright factors. The filled
$F_U$ factors are flat. Expanding each residual factor into flat and bright
parts also shows that there can be no more than $a$ bright factors. Both
conclusions are statements of spectral support and survive arbitrary
superpositions. Applying them to both spins proves
\eqref{sup:graph-spectral}. All jumps annihilate $W$, so both the recycling
and anticommutator terms vanish on every operator supported there. The
assumption $H=0$ then proves stationarity.
\end{proof}

For a $d$-dimensional periodic hypercubic base of side length $n$, let
$M=n^d$ and $p/M\to\nu\in(0,1)$. Fix
$0<\theta<d/(d+1)$. Choose two separated macroscopic strips with
\begin{equation}
 |U_\uparrow|=(1-\theta)p+o(M),\qquad
 |U_\downarrow|=(1-\theta)q+o(M),
 \label{sup:strip-sizes}
\end{equation}
and define the integer $a_\sigma$ by \eqref{sup:region-filling}. Rounding strip
widths changes $O(n^{d-1})$ vertices. If every destination is within radius
$R_n=o(n)$ of its source, discarding the boundary layer gives
\begin{equation}
 |T(U_\sigma)|=d|U_\sigma|-o(M),\qquad
 r(U_\sigma)=O(n^{d-1}).
 \label{sup:source-asymptotics}
\end{equation}
The strict inequality $\theta<d(1-\theta)$ ensures enough source capacity for
all sufficiently large $n$. The two spectral bounds in
\eqref{sup:graph-spectral} squeeze $\Tr(N_b\rho)/M$ to $\theta$, uniformly
over all normalized density matrices supported on $W$. Stirling's formula
applied to \eqref{sup:W-dimension} yields
\begin{equation}
 \frac{\log\dim W}{M}\longrightarrow
 s_d(\theta)=d(1-\theta)
 h_{\rm bin}\!\left(\frac{\theta}{d(1-\theta)}\right)>0,
 \quad h_{\rm bin}(x)=-x\log x-(1-x)\log(1-x).
 \label{sup:entropy-density}
\end{equation}
The logarithm is natural. The result applies to these decorated hypercubic
graphs and the specified transfer class. It does not extend by this proof
to other flat-band geometries with different compact-orbital structure.

\section{Higher-dimensional boundary preparation and overlap cost}
\label{sup:boundary-preparation}

In this section, take independent endpoint cooling,
$\operatorname{Dest}(e)=\{v,w\}$ for $e=(v,w)$. Fill every internal edge
source of each region, so $J_\sigma=E(U_\sigma)$ and
$n_\sigma=|U_\sigma|+|E(U_\sigma)|$. The two regions remain nonadjacent and
their particle numbers sum to $M$. Let $\Psi$ be the normalized dark vector
constructed from these occupied orbitals, and let $\Phi$ be the physical
Fock state occupying every $A_v$ in a region and every internal $B_e$, with
the corresponding spin.

\begin{theorem}[Boundary overlap and local preparation]
\label{sup:boundary-theorem}
If $b_v$ counts edges from $v\in U$ to $U^c$, then
\begin{equation}
 |\langle\Psi|\Phi\rangle|^2
 =c_{\rm Fock}=\prod_\sigma\prod_{v\in\partial U_\sigma}
 (1+\lambda^2b_v)^{-1}.
 \label{sup:area-overlap}
\end{equation}
At bounded maximum degree $\Delta$, a number-conserving circuit of bounded
depth maps $\Phi$ exactly to $\Psi$. Omitting a fixed set $A$ of $k\geq1$
boundary rotations instead produces a non-dark state $\widetilde\Phi$ with
\begin{equation}
 |\langle\Psi|\widetilde\Phi\rangle|^2
 =c_A=\prod_{v\in A}(1+\lambda^2b_v)^{-1}
 \geq(1+\lambda^2\Delta)^{-k}>0.
 \label{sup:dressed-overlap}
\end{equation}
If the witness bound $h_G$ is positive, its subsequent $H=0$ dynamics obeys
$F(t)\leq1-c_A$ and $\Tr[N_b\rho(t)]/M\geq c_Ah_G/M$ for every $t\geq0$.
\end{theorem}
\begin{proof}
All internal $B_e$ modes are already filled. Removing their components from
each occupied $f_v$ leaves
\begin{equation}
 g_v^{\rm raw}=|A_v\rangle
 -\lambda\sum_{\substack{e\ni v\\e\in\partial E(U)}}|B_e\rangle.
 \label{sup:boundary-star}
\end{equation}
A cut edge has only one endpoint in $U$, so these residual columns have
disjoint supports. Their squared norms are $1+\lambda^2b_v$, and they are
orthogonal to the filled internal-edge modes. The normalized all-$A$ Fock
component therefore has exactly the product weight
\eqref{sup:area-overlap}.

For $b_v>0$, define
\begin{equation}
 a_{s_v}^\dagger=b_v^{-1/2}
 \sum_{\substack{e\ni v\\e\in\partial E(U)}}c_{B_e}^\dagger,
 \quad \tan\theta_v=\lambda\sqrt{b_v},\quad
 U_v=\exp[-\theta_v(a_{s_v}^\dagger c_{A_v}
                          -c_{A_v}^\dagger a_{s_v})].
 \label{sup:boundary-gate}
\end{equation}
The spin label is that of the region. This rotation sends the occupied
$A_v$ orbital to $g_v^{\rm raw}/\sqrt{1+\lambda^2b_v}$. The stars have
disjoint supports, also between the two spin regions, and hence all such
gates can act in parallel. If only two-mode rotations are available, each
bounded-size star can be decomposed into at most $\Delta$ such rotations,
so the total depth remains bounded independently of $M$.

Applying all gates gives $\Psi$. Omitting the gates indexed by $A$ leaves
the factors in \eqref{sup:dressed-overlap} unmatched. The factorization of
supports makes their probabilities multiply exactly. For an omitted vertex
$v$, choose a cut edge $e=(v,w)$ with $w\notin U$. The orbital at $A_v$ is
still unrotated and its cut-edge $B_e$ component is absent. An endpoint
transfer $f_w^\dagger d_e$ has nonzero contraction at $A_v$ and creates a
component at the previously empty $A_w$. It therefore acts nontrivially on
this initial state. Finally, Theorem~\ref{sup:graph-theorem} and the witness
monotonicity \eqref{sup:witness-drift} give the all-time bounds.
\end{proof}

For $N_\partial=|\partial U_\uparrow|+|\partial U_\downarrow|$,
\begin{equation}
 (1+\lambda^2\Delta)^{-N_\partial}\leq c_{\rm Fock}
 \leq(1+\lambda^2)^{-N_\partial}.
 \label{sup:area-bounds}
\end{equation}
Thus a boundary with area proportional to $n^{d-1}$ generally incurs an
exponential area cost for this particular bare-Fock overlap. A small overlap
with one witness does not bound the total weight eventually absorbed into
the whole dark subspace. Equation~\eqref{sup:dressed-overlap} instead gives
a constant overlap for the explicitly boundary-prepared family, provided
the number of omitted gates stays fixed.

An integer sequence realizing $p+q=M$ is obtained from two strips of widths
$m_\uparrow,m_\downarrow>2$ that wrap the other $d-1$ periodic directions.
Each strip has
\begin{equation}
 |U_\sigma|=m_\sigma n^{d-1},\qquad
 |E(U_\sigma)|=(dm_\sigma-1)n^{d-1}.
 \label{sup:strip-counts}
\end{equation}
Choose $n+2=(d+1)(m_\uparrow+m_\downarrow)$ and place the strips with empty
gaps between them. Their total particle number is exactly $n^d$. Their width
ratio can approach any desired nonzero spin-density ratio, and
\begin{equation}
 \frac{a_\uparrow+a_\downarrow}{M}
 =\frac d{d+1}-\frac{2}{(d+1)n},\qquad
 \frac{r(U_\uparrow)+r(U_\downarrow)}M\leq\frac4n.
 \label{sup:strip-density}
\end{equation}
With a fixed number of omitted boundary gates, this proves
\begin{equation}
 \liminf_{n\to\infty}\inf_{t\geq0}
 \frac{\Tr[N_b\rho(t)]}{M}\geq c_A\frac d{d+1}>0.
 \label{sup:dressed-thermo}
\end{equation}
On the flat faces, $b_v=1$. Omitting one gate gives
$c_A=(1+\lambda^2)^{-1}$, whereas the unrotated Fock state's overlap with
the same witness is $(1+\lambda^2)^{-4n^{d-1}}$.

\section{Global attraction after bright-occupation dephasing}
\label{sup:recovery}

For this section, assume that $G$ is connected and has minimum degree at
least two. In particular, $B\geq M$. Include the normalized bright-occupation
jumps
\begin{equation}
 Q_{e\sigma}=\bar d_{e\sigma}^\dagger\bar d_{e\sigma},\qquad
 \bar d_e=d_e/\|d_e\|.
 \label{sup:dephasers}
\end{equation}
Each $Q_{e\sigma}$ is a Hermitian projection. The rate of every included
transfer, every incidence selector in both orientations, and every
$Q_{e\sigma}$ is strictly positive and constant. Transfers may use coherent
destination vectors: write them as $a_\sigma^\dagger(z_\mu)d_{e_\mu,\sigma}$,
where $z_\mu\in H_f$ is nonzero. For each spin, their actual destination
vectors must span $H_f$. Independent endpoint transfers satisfy this
condition.

\begin{theorem}[Finite-size global recovery]
\label{sup:recovery-theorem}
Under the preceding assumptions, let $H$ be any Hermitian operator that
preserves the sector \eqref{sup:sector} and satisfies
$H|F_m\rangle=E_F|F_m\rangle$. Then
\begin{equation}
 \Lcal\rho=-i[H,\rho]+\sum_\mu\D[J_\mu]\rho,
 \qquad
 \D[J]\rho=J\rho J^\dagger-\frac12\{J^\dagger J,\rho\},
 \label{sup:repaired-generator}
\end{equation}
where rates are absorbed into the jumps, has the unique stationary state
$\PF$. For every initial density matrix in the sector,
\begin{equation}
 \lim_{t\to\infty}\frac12\|e^{t\Lcal}\rho_0-\PF\|_1=0.
 \label{sup:global-attraction}
\end{equation}
No smallness assumption on $H$ is required.
\end{theorem}

The proof first excludes all nonzero common jump-invariant subspaces
orthogonal to the target. It then uses positivity to show that any wrong
stationary density matrix would produce such a subspace. This second step
is needed to exclude mixed stationary states; common-dark-vector uniqueness
alone would not suffice \cite{Kraus2008Preparation}.

\subsection{The algebra generated by overlapping bright occupations}

Let $\mathcal V\subseteq|F_m\rangle^\perp$ be invariant under every actual
jump. Nonzero jump normalizations and positive rates do not affect this
condition. For one spin, put
$g_{ef}=\langle\bar d_e,\bar d_f\rangle$ and
$E_{ef}=\bar d_e^\dagger\bar d_f$. Fermionic anticommutation gives
\begin{equation}
 [E_{ef},E_{hk}]=g_{fh}E_{ek}-g_{ke}E_{hf}.
 \label{sup:bilinear-commutator}
\end{equation}
In particular,
\begin{align}
 [Q_e,Q_f]&=g_{ef}E_{ef}-g_{fe}E_{fe},\\
 [Q_e,[Q_e,Q_f]]
 &=g_{ef}E_{ef}+g_{fe}E_{fe}-2|g_{ef}|^2Q_e.
 \label{sup:double-commutator}
\end{align}
When two base edges share a vertex, their bright overlap is nonzero. The two
identities therefore isolate both $E_{ef}$ and $E_{fe}$ inside the complex
algebra preserving $\mathcal V$. The edge-overlap graph of a connected base
graph is connected. Repeated use of \eqref{sup:bilinear-commutator} along a
path then gives $E_{ef}$ for every pair. Since the $d_e$ are a basis of
$H_b$, $\mathcal V$ is invariant under all bilinears
$\beta_\alpha^\dagger\beta_\beta$ in any orthonormal bright basis.

The two spin bright numbers $N_{b\sigma}$ belong to this algebra, as do their
finite-spectrum projectors. Hence $\mathcal V$ decomposes into blocks
indexed by $h=(h_\uparrow,h_\downarrow)$. On each block the bright bilinears
generate the full matrix algebra on each exterior-power factor. To see this
without assuming a Slater form for a state, products of bright occupation
projections first isolate a chosen bright Fock configuration. Bilinears then
move its occupied labels to any other configuration of the same particle
number. All configurations must accompany the same flat coefficient vector.
Thus every nonzero block has the form
\begin{equation}
 \mathcal V_h=
 \left(\bigwedge^{h_\uparrow}H_{b,\uparrow}
 \otimes\bigwedge^{h_\downarrow}H_{b,\downarrow}\right)
 \otimes\mathcal M_h,
 \label{sup:invariant-factorization}
\end{equation}
where $\mathcal M_h$ is a subspace of the flat sector with particle numbers
$p-h_\uparrow,q-h_\downarrow$. The factorization is a statement about
invariant subspaces, not about a particular density matrix being a product.

Commuting an actual transfer $a_\sigma^\dagger(z)d_{e\sigma}$ with arbitrary
bright bilinears preserves its destination $z$ and replaces its annihilation
direction by any bright mode. Destination spanning then implies that
$\mathcal V$ is invariant under every operator
\begin{equation}
 a_\sigma^\dagger(z)\beta_\sigma(w),\qquad z\in H_f,\quad w\in H_b.
 \label{sup:all-transfers}
\end{equation}
These derived operators constrain invariant subspaces. The argument does not
claim that every such nonlocal operator is an independently implemented
experimental jump.

\subsection{Selectors can increase both bright numbers}

Decompose the physical annihilator as
$c_{x\sigma}=\alpha_{x\sigma}+\beta_{x\sigma}$ into its flat and bright
parts. The component of a selector that takes bright numbers $h$ to
$h+(1,1)$ is uniquely
\begin{equation}
 \Pi_{h+(1,1)}P_{x\to y}\Pi_h
 =T_{xy,bb}^\dagger\alpha_{x\downarrow}\alpha_{x\uparrow}\Pi_h,
 \label{sup:selector-raise}
\end{equation}
where $T_{xy,bb}^\dagger$ is the creation part with both physical orbitals
projected to $H_b$.

For every physical $x$, an incidence neighbor $y$ can be chosen so that
$\Pi_b|x\rangle$ and $\Pi_b|y\rangle$ are independent. Indeed,
\begin{equation}
 \mathsf D^\dagger|A_v\rangle
 =\lambda\sum_{e\ni v}|e\rangle,\qquad
 \mathsf D^\dagger|B_e\rangle=|e\rangle.
 \label{sup:bright-independence}
\end{equation}
At an incidence pair, minimum degree two prevents these vectors from being
proportional. The restriction of $\mathsf D^\dagger$ to $H_b$ is invertible,
so the same independence holds after physical bright projection.

Suppose a flat vector $v\in\mathcal M_h$ has
$\alpha_{x\downarrow}\alpha_{x\uparrow}v\ne0$. Each species then has at
least one flat particle, and
\begin{equation}
 h_\sigma\leq n_\sigma-1\leq M-2\leq B-2,
 \qquad n_\uparrow=p,\quad n_\downarrow=q.
 \label{sup:bright-capacity}
\end{equation}
Equation~\eqref{sup:invariant-factorization} allows all its existing bright
particles to be placed in the orthogonal complement of the two independent
new bright directions. On that configuration, $T_{xy,bb}^\dagger$ creates a
nonzero two-spin spatial exterior product. Consequently
\eqref{sup:selector-raise} gives a nonzero vector with total bright number
increased by two.

\subsection{Contradiction at maximal bright number}

Choose a nonzero block of $\mathcal V$ whose total bright number
$h_\uparrow+h_\downarrow=h$ is maximal. If $h=M$, all particles are bright
and the flat factor is the vacuum. Repeated application of
\eqref{sup:all-transfers}, starting from arbitrary configurations in the full
bright factors, produces every flat Fock configuration in the $(p,q)$
sector. The span includes $F_m$, contrary to
$\mathcal V\subseteq|F_m\rangle^\perp$.

If $h<M$ and a flat coefficient in $\mathcal M_h$ has a nonzero physical
doublon contraction, \eqref{sup:selector-raise} produces total bright number
$h+2$, contradicting maximality. Therefore every $v\in\mathcal M_h$ obeys
\begin{equation}
 \alpha_{x\downarrow}\alpha_{x\uparrow}v=0
 \quad\text{for all physical }x.
 \label{sup:maximal-no-doublon}
\end{equation}
If $h=0$, Lemma~\ref{sup:flat-target} implies $v\propto F_m$, again a
contradiction.

It remains to consider $0<h<M$. At least one species has a bright particle
while the other has a flat particle. Otherwise, either both species would be
entirely flat or both entirely bright. Relabeling spins if necessary, assume
$h_\uparrow>0$ and $q-h_\downarrow>0$. For a nonzero flat coefficient $v$,
choose a site $x$ such that $\alpha_{x\downarrow}v\ne0$, and set
$z=\Pi_f|x\rangle/\|\Pi_f|x\rangle\|$. The anticommutator
$\eta=\{\alpha_{x\uparrow},a_\uparrow^\dagger(z)\}
=\|\Pi_f|x\rangle\|$ is strictly positive. Using
\eqref{sup:maximal-no-doublon},
\begin{equation}
 \alpha_{x\downarrow}\alpha_{x\uparrow}
 a_\uparrow^\dagger(z)v
 =\eta\alpha_{x\downarrow}v
 +a_\uparrow^\dagger(z)
   \alpha_{x\downarrow}\alpha_{x\uparrow}v
 =\eta\alpha_{x\downarrow}v\ne0.
 \label{sup:created-doublon}
\end{equation}
A transfer \eqref{sup:all-transfers} first reaches total bright number $h-1$
with this flat coefficient. The selector construction then raises both
bright numbers and reaches total $h+1$. This contradicts the original
maximum. All cases are exhausted, proving
\begin{equation}
 \text{there is no nonzero common jump-invariant }
 \mathcal V\subseteq|F_m\rangle^\perp.
 \label{sup:no-wrong-subspace}
\end{equation}
Only the jumps entered this argument; it makes no assumption about the
magnitude of an actual Hamiltonian.

\subsection{Excluding mixed stationary states with a Hamiltonian}

Every jump annihilates $F_m$, and the stated Hamiltonian has $F_m$ as an
eigenvector. In the decomposition
$\mathbb C|F_m\rangle\oplus|F_m\rangle^\perp$, write
\begin{equation}
 J_\mu=\begin{pmatrix}0&A_\mu\\0&B_\mu\end{pmatrix},
 \qquad H=\begin{pmatrix}E_F&0\\0&H_\perp\end{pmatrix}.
 \label{sup:target-blocks}
\end{equation}
Let $\rho$ be stationary, with positive lower block $R$. The target
probability derivative is a sum of nonnegative terms:
\begin{equation}
 0=\Tr[\PF\Lcal\rho]
 =\sum_\mu\Tr(A_\mu R A_\mu^\dagger).
 \label{sup:target-leakage}
\end{equation}
Hence $A_\mu\sqrt R=0$ for every $\mu$. This also implies
$A_\mu^\dagger A_\mu R=R A_\mu^\dagger A_\mu=0$, so the lower stationary
block reduces to
\begin{equation}
 0=-i[H_\perp,R]+\sum_\mu\D[B_\mu]R.
 \label{sup:lower-stationary-block}
\end{equation}
For any $w\perp\supp R$, its diagonal matrix element has vanishing
Hamiltonian and anticommutator contributions because $Rw=0$. Thus
\begin{equation}
 0=\sum_\mu\langle w|B_\mu R B_\mu^\dagger|w\rangle
  =\sum_\mu\|\sqrt R B_\mu^\dagger w\|^2.
 \label{sup:support-invariance}
\end{equation}
Every term vanishes. Therefore $B_\mu\supp R\subseteq\supp R$, while
\eqref{sup:target-leakage} says $A_\mu\supp R=0$. If $R\ne0$, its support
would be a nonzero common jump-invariant subspace in the target orthogonal
complement, forbidden by \eqref{sup:no-wrong-subspace}. Thus $R=0$.
Positivity of $\rho$ then forces its off-diagonal target blocks to vanish,
and trace one gives $\rho=\PF$.

Finally, every limit point of the time-averaged state
$T^{-1}\int_0^T e^{t\Lcal}\rho_0\,dt$ is stationary, because its image under
$\Lcal$ is $(e^{T\Lcal}\rho_0-\rho_0)/T$. Compactness and stationary-state
uniqueness imply convergence of these averages to $\PF$. Meanwhile,
\begin{equation}
 \Lcal^\dagger\PF=\sum_\mu J_\mu^\dagger\PF J_\mu\geq0,
 \label{sup:target-monotone}
\end{equation}
so $F(t)$ is nondecreasing. A bounded nondecreasing function and its Cesaro
average have the same limit; therefore $F(t)\to1$. The pure-state
trace-distance bound
$\frac12\|\rho-\PF\|_1\leq\sqrt{1-\Tr(\PF\rho)}$ proves
\eqref{sup:global-attraction} and completes the theorem.

For a translation-invariant chain with coherent destinations
$z_j=\sum_r\alpha_r f_{j+r}$, destination spanning has a simple finite-size
test. The circulant coefficient matrix is invertible precisely when
\begin{equation}
 A(e^{ik})=\sum_r\alpha_r e^{ikr}\ne0
 \quad\text{for every }k=2\pi m/L.
 \label{sup:coherent-span}
\end{equation}
A zero of this symbol means that the sufficient spanning assumption fails.
It does not by itself prove that the full dynamics has an additional steady
state.

\section{Strong dephasing and the compressed generator}
\label{sup:zeno}

Fix one finite graph and all transfers, selectors, and Hamiltonian
coefficients. Write
\begin{equation}
 \Lcal_\kappa=\kappa\mathcal Q+\mathcal A,\qquad
 \mathcal Q=\sum_{e\sigma}r_{e\sigma}\D[Q_{e\sigma}],
 \qquad r_{e\sigma}>0,
 \label{sup:strong-generator}
\end{equation}
where $\mathcal A$ includes only the fixed transfer and selector dissipators
and the target-preserving Hamiltonian. The parameter $\kappa$ tends to
infinity with everything else, including the graph, held fixed.

\subsection{The conditional expectation selected by dephasing}

In the Hilbert-Schmidt inner product, Hermiticity of each dephaser gives
\begin{equation}
 \langle X,\mathcal QX\rangle_{\rm HS}
 =-\frac12\sum_{e\sigma}r_{e\sigma}
 \|[Q_{e\sigma},X]\|_{\rm HS}^2.
 \label{sup:Q-dirichlet}
\end{equation}
Consequently $\mathcal Q$ is self-adjoint and nonpositive, and its kernel is
the commutant of all $Q_{e\sigma}$. The algebra derived in
Section~\ref{sup:recovery} identifies this commutant explicitly. On a fixed
bright-number block $h=(h_\uparrow,h_\downarrow)$, the bright representation
is the product of two irreducible exterior powers. Different values of
$h_\sigma$ are distinguished by the central phases of $U(B)_\sigma$, so
there are no inter-block intertwiners. The orthogonal kernel projection is
therefore the conditional expectation
\begin{equation}
 \mathcal E(X)=\bigoplus_h
 \frac{I_{b,h}}{d_{b,h}}\otimes\Tr_b(\Pi_hX\Pi_h),
 \qquad d_{b,h}=\binom B{h_\uparrow}\binom B{h_\downarrow}.
 \label{sup:conditional-expectation}
\end{equation}
Only the blocks allowed by the fixed spin particle numbers are included.
Flat coefficient operators in this formula can contain arbitrary many-body
correlations. Equivalently,
\begin{equation}
 \mathcal E(X)=\int_{U(B)_\uparrow\times U(B)_\downarrow}
 U X U^\dagger\,dU,
 \label{sup:haar-expectation}
\end{equation}
where the one-particle bright unitaries are second quantized and act
trivially on flat modes. The measure is normalized Haar measure.

The neighboring $Q_e$ do not generally commute. Strong dephasing therefore
does not preserve a joint occupation pattern of the original overlapping
bright orbitals. It removes their directional information and leaves a
maximally mixed bright factor at each fixed bright particle number.

\subsection{Effective evolution and attraction after projection}

The finite-dimensional strong-dissipation limit gives
\begin{equation}
 e^{t\Lcal_\kappa}=e^{t\Lcal_Z}\mathcal E+O(\kappa^{-1}),
 \qquad
 \Lcal_Z=\mathcal E\mathcal A\mathcal E
 \big|_{\operatorname{ran}\mathcal E},
 \label{sup:zeno-limit}
\end{equation}
uniformly on each fixed interval $0<t_0\leq t\leq T$. The estimate is in any
fixed finite-dimensional operator norm; its constant may depend on the
graph, the rates, $t_0$, and $T$. The exclusion of $t=0$ accommodates an
arbitrary initial state's fast component. This projection principle is an
established strong-coupling result \cite{Burgarth2019StrongCoupling}. It can also be seen by
decomposing the differential equation into $\ker\mathcal Q$ and its
orthogonal complement. On the complement, the spectrum of $\mathcal Q$ is
strictly negative. Integrating the fast equation suppresses its driven
component by $O(\kappa^{-1})$; substituting into the slow equation leaves
$\mathcal E\mathcal A\mathcal E$ at leading order.

\begin{theorem}[Attraction and gap in the strong-dephasing limit]
\label{sup:zeno-theorem}
Under the hypotheses of Theorem~\ref{sup:recovery-theorem}, the compressed
generator $\Lcal_Z$ has the unique stationary density matrix $\PF$ and is
globally attractive on the conditional-expectation state space. Its
finite-size spectral gap is strictly positive, and
\begin{equation}
 \lim_{\kappa\to\infty}g(\Lcal_\kappa)=g(\Lcal_Z)>0,
 \qquad
 g(\mathcal L)=\min_{z\in\operatorname{spec}(\mathcal L)\setminus\{0\}}
 [-\operatorname{Re}z].
 \label{sup:zeno-gap}
\end{equation}
\end{theorem}
\begin{proof}
Define a group-averaged generator on the full operator space by
\begin{equation}
 \overline{\mathcal A}
 =\int\operatorname{Ad}_U\mathcal A\operatorname{Ad}_{U^\dagger}\,dU,
 \qquad \operatorname{Ad}_U(X)=UXU^\dagger.
 \label{sup:averaged-generator}
\end{equation}
It is a legitimate Lindblad generator: its jump terms are the positive
integral of dissipators with jumps $UJ_\mu U^\dagger$, and its Hamiltonian
is $\overline H=\int UHU^\dagger\,dU$. Every $U$ fixes $F_m$ because the
target has no bright particles. Thus all rotated jumps annihilate it and
$\overline H$ retains it as an eigenvector. For any $\rho=\mathcal E\rho$,
\begin{equation}
 \overline{\mathcal A}\rho
 =\int U\mathcal A(\rho)U^\dagger\,dU
 =\mathcal E\mathcal A\rho=\Lcal_Z\rho.
 \label{sup:average-equals-compression}
\end{equation}
The averaged generator is group equivariant and preserves the range of
$\mathcal E$. Its restriction is therefore a completely positive,
trace-preserving evolution on that state space.

Suppose $\rho=\mathcal E\rho$ is stationary for $\Lcal_Z$, and let $R$ be its
target-orthogonal block. Since $\PF$ commutes with the group, $R=\mathcal E R$;
hence $W=\supp R$ is invariant under all bright unitaries and, in particular,
under every $Q_{e\sigma}$. Applying the target-leakage identity to
\eqref{sup:averaged-generator} gives a sum of Haar integrals of continuous
nonnegative functions. A vanishing sum forces each function to vanish at
every group element. Thus every rotated jump has zero leakage from $W$ to
the target. Taking stationary diagonal matrix elements at vectors orthogonal
to $W$ then gives the corresponding nonnegative recycling integrals, exactly
as in \eqref{sup:support-invariance}. Their vanishing forces $W$ to be
invariant under each rotated transfer and selector. In particular, evaluate
at the identity group element: $W$ is invariant under the original transfers
and selectors as well as all the $Q_{e\sigma}$. Equation
\eqref{sup:no-wrong-subspace} excludes a nonzero such $W$ in the target
orthogonal complement. Therefore $R=0$ and $\rho=\PF$.

Target probability is nondecreasing for the averaged evolution. The
time-average argument following \eqref{sup:target-monotone} proves ordinary
trace-norm convergence to $\PF$ within $\operatorname{ran}\mathcal E$.
Density matrices span the Hermitian part of this finite-dimensional
algebra. Convergence on every state therefore excludes nonzero peripheral
eigenvalues on the entire compressed operator space. Boundedness of the
semigroup excludes a Jordan block at zero. Its zero eigenvalue is simple,
and all remaining eigenvalues have strictly negative real parts, proving
$g(\Lcal_Z)>0$.

For the spectral limit, write $\mathcal E_\perp=I-\mathcal E$ and decompose
$\Lcal_\kappa-z$ into its slow and fast blocks. For bounded $z$, the fast
block has an inverse of order $\kappa^{-1}$, and the slow Schur complement is
\begin{align}
 \Lcal_Z-z
 -\mathcal E\mathcal A\mathcal E_\perp
 \bigl[\kappa\mathcal Q_\perp
       +\mathcal E_\perp\mathcal A\mathcal E_\perp-z\bigr]^{-1}
 \mathcal E_\perp\mathcal A\mathcal E.
 \label{sup:schur-complement}
\end{align}
Its last term tends to zero. Thus the bounded eigenvalues converge, with
algebraic multiplicity, to those of $\Lcal_Z$. The remaining eigenvalues
have real parts tending to minus infinity, because they arise from the
strictly negative, self-adjoint spectrum of $\kappa\mathcal Q_\perp$ under
a bounded perturbation. The compressed zero is simple, and
$\Lcal_\kappa$ has an exact simple zero for every $\kappa>0$ by
Theorem~\ref{sup:recovery-theorem}. No additional slow root can approach
zero. The gap limit \eqref{sup:zeno-gap} follows.
\end{proof}

Equation~\eqref{sup:zeno-gap} is a limit at fixed finite size. It supplies no
lower bound uniform in $M$, and it does not imply that the gap increases
monotonically with $\kappa$. The slow spectrum need not be diagonalizable;
although the Schur-complement correction is $O(\kappa^{-1})$, individual
eigenvalues near a defective root need not converge at that same rate.

\subsection{The direct cooling rate survives the projection}

For a fixed block $h$, the conditional expectation makes every normalized
bright one-particle direction have occupation $h_\sigma/B$. Define the
positive destination-frame operator for spin $\sigma$ by
\begin{equation}
 K_{f,\sigma}=\sum_\mu\gamma_\mu\|d_{e_\mu}\|^2
 |z_\mu\rangle\langle z_\mu|>0.
 \label{sup:destination-frame}
\end{equation}
Here $\gamma_\mu$ is the physical rate and $z_\mu$ includes any chosen
destination normalization. If $\rho_f$ is the normalized flat coefficient
state in this block and $n_{f\sigma}=n_\sigma-h_\sigma$, the total cooling
event rate is
\begin{align}
 R_\sigma&=\frac{h_\sigma}{B}\Tr_f
 \left[\bigl(\Tr K_{f,\sigma}
       -c_\sigma^\dagger K_{f,\sigma}c_\sigma\bigr)\rho_f\right]\\
 &\geq\frac{h_\sigma}{B}\lambda_{\min}(K_{f,\sigma})
          (M-n_{f\sigma}).
 \label{sup:averaged-cooling-rate}
\end{align}
The last inequality counts flat holes in an eigenbasis of $K_{f,\sigma}$.
It is strictly positive when $h_\sigma>0$. A general conditional-expectation
state is a convex sum of these block contributions. This calculation shows
why direct dissipative transfers remain active after the strong-dephasing
projection. It does not make total bright number a Lyapunov function:
selectors can create bright particles.

\section{Finite-time upper and lower bounds}
\label{sup:times}

\subsection{A constructive finite-word convergence certificate}

The invariant-subspace result supplies a finite-size rate certificate
without assuming an unproved estimate for the spectral gap. It can be very
conservative. The fixed-sector Hilbert dimension is
\begin{equation}
 \mathscr D=\binom{M+B}{p}\binom{M+B}{q}.
 \label{sup:hilbert-dimension}
\end{equation}
Choose a reference frequency $\Gamma>0$, set $s=\Gamma t$,
$\widehat J_\mu=J_\mu/\sqrt\Gamma$, $\widehat H=H/\Gamma$, and define
\begin{equation}
 K=-i\widehat H-\frac12\sum_\mu\widehat J_\mu^\dagger\widehat J_\mu.
 \label{sup:no-jump-K}
\end{equation}
These are dimensionless quantities.

Starting with $R_0=\operatorname{span}\{F_m\}$, form
\begin{equation}
 R_{n+1}=R_n+\sum_\mu\widehat J_\mu^\dagger R_n.
 \label{sup:reverse-reachability}
\end{equation}
If an incomplete subspace stabilized, its nonzero orthogonal complement
would be jump invariant and orthogonal to the target, contradicting
\eqref{sup:no-wrong-subspace}. Its dimension therefore increases until it is
full. For some $m\leq\mathscr D-1$, $R_m$ is the whole sector. Equivalently,
the finite word Gram matrix
\begin{equation}
 G_m=\sum_{n=0}^m\sum_{|w|=n}w^\dagger\PF w>0,
 \qquad w=\widehat J_{\mu_n}\cdots\widehat J_{\mu_1},
 \qquad g=\lambda_{\min}(G_m)>0,
 \label{sup:word-gram}
\end{equation}
is strictly positive; the empty word is the identity. It can be accumulated
without storing individual words through
$G^{(0)}=\PF$,
$G^{(n+1)}=\sum_\mu\widehat J_\mu^\dagger G^{(n)}\widehat J_\mu$,
and $G_m=\sum_{n=0}^mG^{(n)}$.

\begin{theorem}[Finite-size exponential certificate]
\label{sup:word-theorem}
Under the recovery assumptions, define
\begin{align}
 S_J&=\sum_\mu\|\widehat J_\mu\|^2,
 &B_m&=\sum_{n=0}^mS_J^n,& k&=\|K\|,\\
 \tau&=\min\left\{1,
 \frac{\log[1+\sqrt g/(2\sqrt{B_m})]}k\right\},
 &\eta&=\min\left\{\frac12,\frac{g\tau^m}{4m!}\right\}.
 \label{sup:certificate-constants}
\end{align}
For $k=0$, the second term in the minimum defining $\tau$ is interpreted as
infinite. Then $\tau,\eta>0$, and every initial state satisfies
\begin{align}
 1-F(t)&\leq(1-\eta)^{\lfloor\Gamma t/\tau\rfloor}[1-F(0)],
 \label{sup:certificate-fidelity}\\
 \frac12\|\rho(t)-\PF\|_1
 &\leq\sqrt{1-F(0)}
 (1-\eta)^{\lfloor\Gamma t/\tau\rfloor/2}.
 \label{sup:certificate-trace}
\end{align}
In particular, worst-case trace-distance accuracy $\varepsilon\in(0,1)$ is
reached no later than
\begin{equation}
 t_\varepsilon\leq\frac\tau\Gamma
 \left\lceil\frac{2\log(1/\varepsilon)}{-\log(1-\eta)}\right\rceil.
 \label{sup:certificate-time}
\end{equation}
\end{theorem}
\begin{proof}
For a word of length $n$, the trajectory Kraus operator over an interval
$\tau$ has the form
\begin{equation}
 V_w(\mathbf s)=e^{Ks_{n+1}}\widehat J_{\mu_n}e^{Ks_n}\cdots
 \widehat J_{\mu_1}e^{Ks_1},\qquad
 s_i\geq0,\quad\sum_{i=1}^{n+1}s_i=\tau.
 \label{sup:trajectory-word}
\end{equation}
Telescoping the differences obtained by replacing the exponential factors
one at a time by the identity, and using $\|e^{Ku}\|\leq e^{ku}$, gives
\begin{equation}
 \|V_w(\mathbf s)-w\|
 \leq(e^{k\tau}-1)\prod_{j=1}^n\|\widehat J_{\mu_j}\|.
 \label{sup:word-error}
\end{equation}
For a unit vector $\psi$, apply
$|a+b|^2\geq|a|^2/2-|b|^2$ to the amplitude
$\langle F_m|V_w|\psi\rangle$. First average over the normalized
time-ordering simplex for each word, then sum over all words of lengths
$0\leq n\leq m$. The resulting positive-operator sum is bounded below by
\begin{equation}
 \frac12G_m-(e^{k\tau}-1)^2B_m I\geq\frac g4I.
 \label{sup:averaged-word-bound}
\end{equation}
The last step is the definition of $\tau$.

In the full trajectory expansion, the actual time-ordering simplex has
volume $\tau^n/n!$. Since $\tau\leq1$, each such volume with $n\leq m$ is
at least $\tau^m/m!$. Every trajectory contribution to the adjoint target
effect is positive, including the terms with more than $m$ jumps. Keeping
only the first $m$ orders and using \eqref{sup:averaged-word-bound} therefore
gives
\begin{equation}
 E_F(\tau):=e^{\tau\widehat{\Lcal}^\dagger}\PF
 \geq\frac{g\tau^m}{4m!}I.
 \label{sup:target-effect-lower}
\end{equation}
The operator $E_F(\tau)$ is an effect, so $0\leq E_F(\tau)\leq I$.
The target is stationary, hence its expectation in $F_m$ is exactly one.
Positivity of $I-E_F$ then forces its target off-diagonal blocks to vanish.
Combining this with \eqref{sup:target-effect-lower} gives
\begin{equation}
 E_F(\tau)\geq\PF+\eta(I-\PF).
 \label{sup:one-window-absorption}
\end{equation}
Each time window of physical length $\tau/\Gamma$ consequently reduces the
failure probability by at least a factor $1-\eta$. Iteration, followed by
target monotonicity for the remaining fractional window, proves
Eq.~\eqref{sup:certificate-fidelity}. The pure-target trace-distance
inequality proves Eq.~\eqref{sup:certificate-trace} and
Eq.~\eqref{sup:certificate-time}.
\end{proof}

The constants use only a finite jump-word Gram matrix and operator norms;
they are not defined implicitly through an unknown gap. Their dependence on
the Hilbert dimension can nevertheless be severe. This bound proves a finite
preparation time at each finite size, not a polynomial upper bound in $M$.
For computation, a less conservative alternative is a rigorously controlled
positive lower bound on
\begin{equation}
 \eta_{\rm num}(T)=\lambda_{\min}
 \left[(I-\PF)e^{T\Lcal^\dagger}\PF(I-\PF)\right]_{F_m^\perp}.
 \label{sup:numerical-effect}
\end{equation}
A floating-point value close to zero without an error bound is a numerical
cross-check, not a rigorous certificate.

\subsection{A size-independent weak-perturbation window for chain witnesses}

Use the normalized chain convention of Section~\ref{sup:chain} and consider
\begin{equation}
 \Lcal=\Lcal_0+\kappa\sum_{j\sigma}\D[Q_{j\sigma}]
 -i\epsilon_H[H_{\rm MT},\,\cdot\,],\qquad
 H_{\rm MT}=4\sum_{j\sigma}Q_{j\sigma}+U\sum_xq_x.
 \label{sup:weak-generator}
\end{equation}
Here $\kappa\geq0$, the actual Hamiltonian coefficient is $\epsilon_H$, and
$Q_j=d_j^\dagger d_j$ is a projection because the chain $d_j$ is normalized.
Let $P_\Psi$ be the dark witness in \eqref{sup:chain-witness}.

For any normalized pure vector $\Psi$ and any projection $Q$, decompose
$\Psi$ into the $Q=0$ and $Q=1$ sectors. The two nonzero eigenvalues of
$\D[Q]P_\Psi$ are
$\pm\sqrt{\operatorname{Var}_\Psi Q}/2$: the dissipator retains just minus
one half of the two off-diagonal blocks between these sectors. Similarly,
decomposing $H\Psi$ into its component parallel to $\Psi$ and an orthogonal
residual shows that the nonzero eigenvalues of $i[H,P_\Psi]$ are
$\pm\Delta_\Psi H$. Since \eqref{sup:witness-drift} remains nonnegative for
$\Lcal_0$, the witness probability $w(t)=\Tr[P_\Psi\rho(t)]$ obeys
\begin{equation}
 \dot w(t)\geq-\kappa A_S-|\epsilon_H|\Delta_\Psi H_{\rm MT},
 \qquad A_S=\frac12\sum_{j\sigma}
             \sqrt{\operatorname{Var}_\Psi Q_{j\sigma}}.
 \label{sup:weak-drift}
\end{equation}

These variances are confined to the boundary of the witness. For one spin
interval $I=[A,B]$, only $Q_j$ with $j\in[A-1,B]$ can be nonzero. For
$j\in[\min J+1,\max J-1]$, all three physical modes in $d_j$ are fully
occupied by that spin, so $Q_j\Psi=\Psi$. Outside the interval support,
$Q_j\Psi=0$. Therefore at most $w+3$ dephasers per spin have nonzero
variance. Since a projection has variance at most $1/4$,
\begin{equation}
 A_S\leq\frac{w+3}{2}.
 \label{sup:variance-count}
\end{equation}
The interaction term annihilates $\Psi$. Applying the triangle inequality
to $(H_{\rm MT}-\langle H_{\rm MT}\rangle)\Psi$, rather than using the
extensive norm of the Hamiltonian, gives
\begin{equation}
 \Delta_\Psi H_{\rm MT}
 \leq4\sum_{j\sigma}\sqrt{\operatorname{Var}_\Psi Q_{j\sigma}}
 \leq4(w+3).
 \label{sup:H-variance}
\end{equation}
Integration from the Fock initial state with $w(0)=c_S$ proves
\begin{align}
 w(t)&\geq w_{\rm low}(t)
 =\left[c_S-(w+3)(\kappa/2+4|\epsilon_H|)t\right]_+,\\
 F(t)&\leq1-w_{\rm low}(t),\qquad
 \frac{\Tr[N_b\rho(t)]}{L}
 \geq\left(\frac12-\frac{w+4}{L}\right)w_{\rm low}(t).
 \label{sup:weak-window}
\end{align}
The last two inequalities use the fixed operator bounds
\eqref{sup:witness-operator}, which remain valid even though $\Psi$ is no
longer stationary. The permanent obstruction becomes a finite-time
retention bound.

When $\epsilon_H=0$, achieving $F\geq1-\varepsilon_F$ with
$\varepsilon_F<c_S$ requires
\begin{equation}
 t\geq\frac{2(c_S-\varepsilon_F)}{(w+3)\kappa}.
 \label{sup:weak-time-lower}
\end{equation}
This is an initial-state preparation-time lower bound of order
$\kappa^{-1}$. It does not determine a perturbative law for the Liouvillian
gap. If the chain transfer family also satisfies the destination-spanning
condition, every fixed $\kappa>0$ prepares the target at each finite $L$.
Continuity of the finite-time semigroup in $\kappa$, together with
\eqref{sup:chain-bounds}, then gives the noncommuting limits
\begin{equation}
 \lim_{\kappa\downarrow0}\lim_{t\to\infty}F_\kappa(t)=1,
 \qquad
 \lim_{t\to\infty}\lim_{\kappa\downarrow0}F_\kappa(t)
 =F_0(\infty)\leq1-c_S.
 \label{sup:noncommuting-limits}
\end{equation}
The second limit exists because target probability is monotone. This is a
finite-system order-of-limits statement.

\subsection{A transport bound for locally conserved particle number}

Consider a bounded-degree graph family with uniformly bounded interaction
range and support size. Decompose the actual generator into local Hermitian
terms $H_Z$ and local jumps $J_\mu$ supported on $Z$. Assume each term
conserves the total particle number on its own support. For nonnegative
weights $w_x$, define
\begin{equation}
 X_w=\sum_xw_xn_x,\qquad n_x=n_{x\uparrow}+n_{x\downarrow},
 \qquad
 \operatorname{osc}_Z(w)=\max_{x\in Z}w_x-\min_{x\in Z}w_x.
 \label{sup:transport-observable}
\end{equation}
In a commutator with an operator supported on $Z$, a common constant can be
subtracted from every $w_x$ in that support. Since $\|n_x\|=2$, the elementary
commutator estimate gives
\begin{equation}
 \|[X_w,J_\mu]\|
 \leq4|Z|\operatorname{osc}_Z(w)\|J_\mu\|.
 \label{sup:local-commutator}
\end{equation}
Using
$\D[J]^\dagger X=(J^\dagger[X,J]+[J^\dagger,X]J)/2$ and the analogous
Hamiltonian estimate yields
\begin{equation}
 \left|\frac{d}{dt}\langle X_w\rangle_t\right|\leq C_w,
 \qquad
 C_w=4\sum_Z|Z|\operatorname{osc}_Z(w)
 \left(\|H_Z\|+\sum_{\mu:\supp J_\mu=Z}\|J_\mu\|^2\right).
 \label{sup:transport-constant}
\end{equation}
Because $X_w/\|X_w\|$ is an effect, trace-distance accuracy
$\frac12\|\rho(t)-\PF\|_1\leq\varepsilon$ implies
$|\langle X_w\rangle_t-\langle X_w\rangle_F|
\leq\varepsilon\|X_w\|$. Integrating \eqref{sup:transport-constant} gives
the finite-size lower bound
\begin{equation}
 t\geq\frac{
 [\langle X_w\rangle_F-\langle X_w\rangle_0
       -\varepsilon\|X_w\|]_+}{C_w}.
 \label{sup:transport-time}
\end{equation}

For the macroscopically empty-region initial states above, choose $w_x$ as
the graph distance to their full physical support. In the boundary-prepared
case, this support includes every cut-edge site appearing in a rotation.
Then $\langle X_w\rangle_0=0$. In the target, the spin-summed one-body density
matrix is $\Pi_f$: the fully polarized state fills the flat space, and a
global spin lowering does not change a spin-scalar one-body expectation.
Thus
\begin{equation}
 \langle n_{A_v}\rangle_F=(G_f^{-1})_{vv}
 \geq\frac1{1+2\lambda^2\Delta}.
 \label{sup:target-density-lower}
\end{equation}
The inequality follows from
$\|\mathsf B\mathsf B^{\mathsf T}\|\leq2\Delta$.
On a $d$-dimensional system of linear size $n$, a positive fraction of the
empty-region vertices are a distance proportional to $n$ from the initial
support. Hence $\langle X_w\rangle_F\geq c_1Mn$ and
$\|X_w\|\leq c_2Mn$, for positive constants fixed by the geometry. A fixed
number of bounded-strength, bounded-range channels per site gives
$C_w\leq c_3M$. For any fixed accuracy
$0<\varepsilon<c_1/c_2$, \eqref{sup:transport-time} therefore implies
\begin{equation}
 t_\varepsilon\geq c_4n=c_4M^{1/d},\qquad c_4>0.
 \label{sup:linear-size-time}
\end{equation}
This bound applies to the repaired protocol as well. It is a consequence
of local particle transport, consistent with locality bounds for Markovian
dynamics \cite{Poulin2010Locality,Barthel2012Quasilocality}; no diffusive scaling assumption is
used. If rates, including $\kappa$, grow with system size, the constant in
this fixed-strength bound also changes.

\section{Numerical methods and independent checks}
\label{sec:numerics}

\subsection{Conventions, bases, and observables}

All numerical examples use $\lambda=1/\sqrt{2}$. A cooling channel is
$\bar f_{v\sigma}^{\dagger}\bar d_{e\sigma}$, where both one-particle
orbitals have unit physical norm. Each included cooling channel has rate
$\gamma=1$, each directed incidence selector has rate $\eta=1$, and each
normalized bright-occupation projector has rate $\kappa$. The three-cell
example has two independent endpoint destinations for every edge and spin:
there are $12$ cooling, $12$ selector, and $6$ dephasing channels. The
four-cell example has the single destination $e=j\mapsto v=j$, giving
$8$ cooling, $16$ selector, and $8$ dephasing channels. Speeds across these
two protocols are not compared, since their local channel resources differ.

When present, the Hamiltonian is
\begin{equation}
 H=hH_{\rm par},\qquad
 H_{\rm par}=\sum_{e,\sigma}d_{e\sigma}^{\dagger}d_{e\sigma}
       +1.7\sum_x n_{x\uparrow}n_{x\downarrow}.
\end{equation}
The $d$ orbitals in this expression are \emph{unnormalized}. This convention
must not be substituted into a formula written for the normalized-chain
Hamiltonian without changing its coefficients.

We enumerate the physical Fock basis with all spin-up modes preceding all
spin-down modes. Fermionic signs are evaluated from the occupied modes
below the annihilated mode. The many-body Hilbert dimensions are
\begin{equation}
 D=\binom{M+B}{p}\binom{M+B}{q}
 =90\quad(M=3,p=2,q=1),\qquad
 D=784\quad(M=4,p=q=2).
\end{equation}
For column vectorization, the sparse matrix is
\begin{align}
 \mathbb L={}&-i(I\otimes H-H^{\mathsf T}\otimes I)\nonumber\\
 &+\sum_\mu\left[
 \overline{J_\mu}\otimes J_\mu
 -\frac12 I\otimes J_\mu^\dagger J_\mu
 -\frac12(J_\mu^\dagger J_\mu)^{\mathsf T}\otimes I\right].
\end{align}
The jumps in this formula include square roots of their rates. Complex
conjugation and transpose are kept distinct.

The target is first constructed in an orthonormal flat/bright Fock basis
and transformed to the physical basis by exterior powers of the one-particle
orthogonal transformation. The bright number uses the physical orthogonal
projector onto the columns of $\mathsf D$, obtained by QR decomposition.
It is not the source count. We save $F(t)$, $\langle N_b\rangle_t/M$, the
doublon density per physical site, trace, Hermiticity error, and
$\|\mathbb L\operatorname{vec}\rho(t)\|_2$.

\subsection{Static checks at larger sizes}

For the chain, we evaluate the four exit sets
$\{0\}$, $\{0,1\}$, $\{-1,0,1\}$, and $\{-1,2\}$.
For each set, $b=3,5,10,20,40,80$ and $a=b$ or $a=2b$.
The definitions $p=2a+w$, $q=2b+w$, and $L=2(a+b+w)$ therefore give
$48$ instances with $L\leq486$, all within the stated theorem sequence.
For each spin, the occupied physical-orbital matrix is
$Z=[\mathsf F_I,U_J]$. An economy QR factorization gives its occupied
orthonormal frame. The Fock overlap is the squared determinant of the
submatrix on the specified occupied physical sites, evaluated through a
logarithmic determinant. This calculation does not use the analytic
$D_n$ recurrence. The maximum difference from $c_S$ is
$1.38\times10^{-14}$.

Sparse solves with $G_f$ apply $\Pi_f=\mathsf F G_f^{-1}\mathsf F^\dagger$
without constructing a system-sized dense projector. After removing the
already occupied flat directions, the eigenvalues of the residual flat
projection test its rank. The largest absolute eigenvalue beyond the two
allowed directions is $7.66\times10^{-17}$; the largest QR orthogonality
residual is $1.47\times10^{-14}$. The plotted bright densities are direct
one-particle projector expectations in the occupied Slater spaces. The
dashed curves are spectral lower bounds, not predicted steady-state
densities.

For the two-dimensional entropy check, take even $n$ divisible by four,
$p=q=n^2/2$, and two separated strips of width $m=n/4$. Each strip contains
$mn$ vertices and $(2m-1)n$ internal edges, and the source number is
$a=n^2/2-mn=n^2/4$. We evaluate the exact combinatorial dimension with
logarithmic gamma functions for
$n=12,20,32,48,64,96,128,256$. Thus
\begin{equation}
 \frac{\log\dim W}{n^2}
 =\frac{2}{n^2}\log\binom{(2m-1)n}{n^2/4}
 \longrightarrow\log2.
\end{equation}
The smaller $n=12,20,32$ cases also have independent occupied-space QR
and bright-projector calculations. The larger entries are geometric and
combinatorial checks, not many-body simulations.

For the area law, the balanced strips have
$n=2(d+1)m-2$ and $m=3,4,5,8,12,20$. Their total particle number is
exactly $n^d$. At $d=2$ and $m=3,4,5$ ($n=16,22,28$), physical-orbital
QR reproduces the logarithm of the bare Fock overlap
$-4n\log(3/2)$ to better than $10^{-10}$. Independently constructed
boundary-star orbitals reproduce the probabilities $2/3$ and $4/9$ when
one or two actual boundary rotations are omitted. The larger two-dimensional
and all three-dimensional entries evaluate the exact geometry and local
gate factors only. They are not reported as dense projector checks or
full initial-state dynamics.

\subsection{A finite example with an actual wrong dark state}

The three-cell endpoint protocol already has a one-dimensional common
dark kernel at $\kappa=0$. Its trajectories therefore test finite-graph
attraction and changes of rate, rather than the removal of an actual
wrong dark state. A separate four-cell single-exit example supplies the
latter test. On the periodic chain with labels $0,1,2,3$, define
\begin{equation}
 |\Psi\rangle\ \propto\
 f_{0\uparrow}^{\dagger}u_{0\uparrow}^{\dagger}
 f_{2\downarrow}^{\dagger}u_{2\downarrow}^{\dagger}|0\rangle,
 \qquad
 |\chi\rangle=
 \frac{(I-P_F)|\Psi\rangle}{\|(I-P_F)|\Psi\rangle\|}.
\end{equation}
Each occupied source has its single flat destination occupied, and the
physical supports of the two spins are disjoint. Consequently $\chi$
is a dark state orthogonal to the target. Its largest cooling/selector
residual is $4.84\times10^{-17}$, its target overlap is below
$5.56\times10^{-17}$, and
$\langle\chi|N_b|\chi\rangle/M=0.3913043478$.
Direct checks of the full sparse cooling/selector and dephasing
generators give trace-operator residuals below $6.29\times10^{-15}$
and target-density residuals below $3.40\times10^{-16}$.
The unrepaired density residual of $|\chi\rangle\langle\chi|$ is
$2.11\times10^{-16}$, whereas the unit-rate dephasing contribution
has residual $0.5639532652$.
The physical Fock initial state has up spins on $A_0,B_0$ and down spins
on $A_2,B_2$. Its squared overlap with $\chi$ is $0.4410153554$.
The dark-weight argument therefore gives the finite-example bound
\begin{equation}
 F(t)\leq1-|\langle\chi|\Phi\rangle|^2
       \simeq0.5589846446\qquad(\kappa=0,H=0)
\end{equation}
for this Fock state. This bound is separate from the macroscopic sequence:
$M=4$ does not satisfy $a,b\geq3$, and we do not infer a positive
macroscopic bright-density bound from this small system.

Sparse matrix-exponential actions propagate both initial states to
$\gamma t=100$ for $\kappa=0,1$. For the wrong dark initial state,
$F(100)=0.7275927379$ at $\kappa=1$, while the $\kappa=0$ trajectory
remains stationary to numerical precision. For the physical Fock state,
the corresponding values are $0.7384524102$ and $0.2429181505$.
These are finite-time values, not assertions of saturation at $t=100$.

For longer trajectories, we use an exact translation reduction. The
generator and all three reported observables commute with translations.
Replacing the initial density by its translation average therefore leaves
their expectation values unchanged at every time. An orthonormal
operator-space embedding reduces the $614656$ Liouville coordinates to
$153728$ translation-invariant coordinates. Its orthogonality residual is
$2.52\times10^{-15}$, and deterministic random-vector checks of generator
invariance give relative residuals below $5.56\times10^{-16}$.
The reconstructed density in this calculation is the translation-averaged
state; we do not identify it with the original pure-state density.
For $\chi$ at $\kappa=1$, the extended trajectory reaches
$F(500)=0.9990602765$ and
$\langle N_b\rangle_{500}/M=1.303585010\times10^{-4}$.
Its invariant observables agree with the unreduced evolution at shared
times through $t=100$ within $1.78\times10^{-15}$.
For the Fock state at $\kappa=0$, the extended values are
$F(500)=0.3998700872$ and
$\langle N_b\rangle_{500}/M=0.2335721344$.
At $\kappa=1$, the same Fock state instead reaches
$F(500)=0.9990977392$ and
$\langle N_b\rangle_{500}/M=1.251616705\times10^{-4}$.
Across all three extended runs, the largest trace error is
$2.19\times10^{-13}$ and the largest Hermiticity error is
$8.30\times10^{-16}$. Density eigenvalues checked at $t=0,250,500$
are no smaller than $-1.60\times10^{-16}$. The largest discrepancy
from the full-density invariant observables at shared times through
$t=100$ is $2.73\times10^{-15}$.
These finite-time residuals are consistent with unblocking for positive
measurement rate and persistent obstruction without it.

\subsection{Three-cell dynamics, slow spectra, and numerical tolerances}

For $M=3$, we propagate a target initial state and the physical Fock state
with up spins on $A_0,A_1$ and a down spin on $A_2$. We use
$h=0,0.6$ and $\kappa=0,0.1,1,10$, with $251$ equally spaced times
through $\gamma t=250$ for the Fock state and six times for the target.
The matrix-exponential action uses SciPy's double-precision adaptive
Taylor implementation. No density renormalization or eigenvalue clipping
is applied. Across all $16$ runs, the maximum trace error is
$6.98\times10^{-13}$, the maximum Frobenius Hermiticity error is
$3.46\times10^{-14}$, and the smallest density eigenvalue is no less
than $-5.29\times10^{-16}$. Positivity is checked at every saved time.
For the four-cell full-density runs, the corresponding trace and
Hermiticity errors are at most $4.05\times10^{-14}$ and
$3.24\times10^{-16}$. Their density eigenvalues are checked at
$t=0,50,100$, with minimum $-8.63\times10^{-16}$.

The three-cell Liouvillian splits into three translation sectors of
dimension $2700$. We compute the $16$ eigenvalues with largest real part
in each sector, using Arnoldi tolerance $2\times10^{-11}$ and Krylov
dimension at least $64$. We include $\kappa=0$ as a control and
$\kappa=0.1,0.3,1,3,10,30,100,300$ for the strong-dephasing scan,
at both Hamiltonian strengths. This gives $54$ sector calculations.
The maximum eigenpair residual is $7.97\times10^{-12}$. Each parameter
point resolves one zero eigenvalue across the three sectors, using
$|z|<10^{-8}$ as the zero threshold. The numerical gap is determined by
the largest \emph{nonzero real part}, not the smallest eigenvalue modulus.
A shift-invert check at $h=0.6,\kappa=300$ agrees with the gap from
largest-real-part Arnoldi within $3.7\times10^{-12}$.

The Zeno generator is formed only on the $190$-dimensional image of the
band-Fock conditional expectation. Its embedded basis has unit
Hilbert-Schmidt norm, and we evaluate $V^\dagger\mathbb A V$.
This avoids the artificial zero eigenvalues of an uncompressed
$\mathcal E\mathcal A\mathcal E$ in the full operator space.
The compression has a single zero root and
\begin{equation}
 g_Z=0.0778747772931\quad(h=0),\qquad
 g_Z=0.0775354346177\quad(h=0.6).
\end{equation}
At $\kappa=300$, the full gaps are $0.0773602730742$ and
$0.0771691028571$, respectively. They are $0.661\%$ and $0.472\%$
below their own fixed-size limits. These computations test the
fixed-$M$ convergence statement; they establish neither a
thermodynamic gap nor a dynamical scaling exponent.

\begin{table}[htbp]
\centering
\caption{Three-cell gaps under the same per-channel normalization.
The last row is the conditional-expectation compression.}
\begin{tabular}{ccc}
\hline
$\kappa$ & $g$, $h=0$ & $g$, $h=0.6$\\
\hline
0 & 0.0409644953 & 0.0429633343\\
0.1 & 0.0419966853 & 0.0436650211\\
0.3 & 0.0438661188 & 0.0450309160\\
1 & 0.0489668196 & 0.0493463892\\
3 & 0.0573610028 & 0.0579312006\\
10 & 0.0674405392 & 0.0687645790\\
30 & 0.0734393642 & 0.0742026830\\
100 & 0.0763867465 & 0.0764659996\\
300 & 0.0773602731 & 0.0771691029\\
$\infty$ & 0.0778747773 & 0.0775354346\\
\hline
\end{tabular}
\end{table}

The finite-graph checks and length-three word-Gram calculation
were rerun from copied source in a separate output directory. The word
certificate uses $h=0$, $\kappa=1$, and reference frequency $\Gamma=1$,
with all other per-channel rates as specified above. These checks
reproduce the $1,22,89,90$ reverse-reachable dimensions, the nine
two-dimensional boundary checks, and the conservative constants
$g=0.0872398358217$, $\tau=1.47001518196\times10^{-4}$, and
$\eta=1.15469971352\times10^{-14}$. These are floating-point evaluations
of the proof's construction, not interval-arithmetic certificates or
estimates of a typical mixing time. Earlier historical validation runs
were not repeated and are not counted among these results.

\begin{figure}[htbp]
\centering
\includegraphics[width=\textwidth]{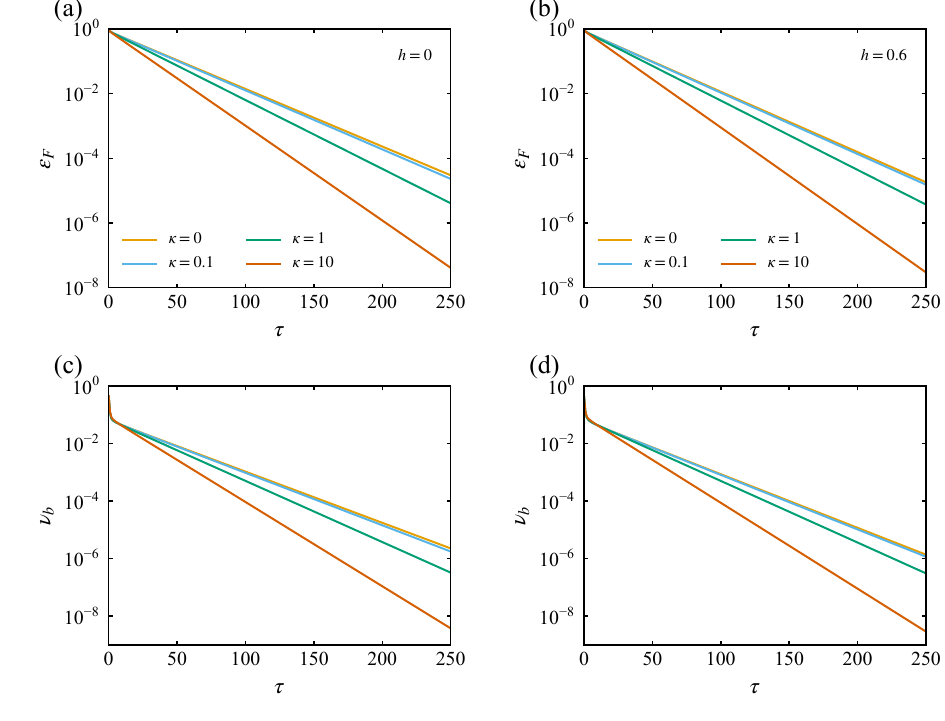}
\caption{Three-cell endpoint-cooling controls for a physical Fock initial
state: up spins occupy $A_0,A_1$ and the down spin occupies $A_2$.
Here $\tau=\gamma t$, $\varepsilon_F=1-F$, and
$\nu_b=\langle N_b\rangle/M$. The top row shows infidelity
$\varepsilon_F$ and the bottom row shows physical bright density $\nu_b$,
for $h=0$ (left) and $h=0.6$ (right).
All other normalizations are identical. The unrepaired three-cell
protocol is already attractive in this sector, so these panels do not
demonstrate the removal of a wrong dark state.}
\label{fig:three-cell-controls}
\end{figure}

\clearpage
\bibliographystyle{unsrtnat}
\bibliography{references}